\documentclass[a4paper,12pt]{article}
\usepackage{epsfig}
\usepackage{graphicx}
\usepackage{rotating}
\usepackage{amsmath}
\usepackage{bm}
\usepackage[letterpaper, left=3cm, right=3cm, top=3cm]{geometry}

\usepackage[svgnames,table]{xcolor} % Colored tables package
\usepackage{lscape} %Adds landscape mode for appendix tables
\usepackage{booktabs} % Nicer tables package
\usepackage{pifont} %Checkmarks in tables
\usepackage{dirtytalk} % for proper quotes
\usepackage[normalem]{ulem} %for strikethrough text

\usepackage{placeins}
\usepackage{pdfpages}

\usepackage{url}

\usepackage{dcolumn}

\usepackage[style=apa]{biblatex}
\begin{document}

\title{\vspace{0cm} Mechanisms for belief elicitation without ground truth

}

\author{Niklas V. Lehmann}

\date{\textit{May 8, 2025}}
\maketitle

\begin{abstract} 
This review article examines the challenge of eliciting truthful information from multiple individuals when such information cannot be verified, a problem known as \say{information elicitation without verification}. This article reviews over 25 mechanisms designed to incentivize truth-telling in such scenarios and their effectiveness in empirical studies. Although many mechanisms theoretically ensure truthfulness as a Bayesian Nash Equilibrium, empirical evidence regarding the effects of mechanisms on truth-telling is limited and generally weak. Consequently, more empirical research is needed to validate mechanisms. However, empirical validation is difficult because most mechanisms are very complex and cannot be easily conveyed to research subjects. This review suggests that simple and intuitive mechanisms may be easier to empirically test and apply.

\noindent
\end{abstract}
\vspace{0.5cm} \noindent

\vspace{6pt}


\thispagestyle{empty}
\clearpage
\addtocounter{page}{-1}

\setlength{\baselineskip}{24pt}

%-----------------------------------------------%

% ----------------------------------------------- %
% Nomenclature: 

% Forecasters/Raters/workers/players/experts/subjects = respondents

% Principal/center/principal = principal

% Forecasts/reports/Submitted beliefs = reports

% Games/ Mechanisms / Methods = mechanisms

% Incentivise or Incentivize?

%Amazon Mechanical Turk = MTurk

%truthful/honest/truthtelling = truthful

% ----------------------------------------------- %

\clearpage
\setcounter{page}{1}

%---------------------------------------1 Introduction ---------------------------------------------------------------%
\section{Introduction}

%\linenumbers

%|% Background: Prelec comes up with the BTS to to solve situations in which "objective truth is unknowable" 
\textcite{prelec_bayesian_2004} introduced the \say{Bayesian Truth Serum} (BTS), a mechanism to \say{elicit truthful subjective data in situations where objective truth is unknowable} by creating incentives for truth-telling. 
%|% Background: 
%|% Given that these mechanisms could aid us in many important areas, such as e.g. long-term forecasting, estimating risks and the effects of interventions, making sure self-reported data is correct and many more. Why are these mechanisms not commonplace in practice? How well do they work? How can we get them into application if they do work well?
The BTS may aid us in many important areas, such as long-term forecasting\footnote{The research agenda of the Global Priorities Institute specifically calls for research in that direction \parencite{greaves_research_2020}.}, estimating risk, data labeling, and ultimately enhancing the accuracy of any self-reported data - at least this is what \textcite{prelec_bayesian_2004} suggests. 
%|% The actual problem: There are a ton of different mechanisms up for choice.
Since 2004, numerous mechanisms for \textit{information elicitation without verification} (IEWV) have been proposed. It has become increasingly difficult for researchers and practitioners to decide which mechanism to use, or even whether to use one at all.
%|% What this study did: Document the state of the art, collect evidence and help to better understand the problem at the heart of the subject matter.
To help researchers and practitioners gain oversight over a highly fragmented literature, this article examines over 25 existing mechanisms and reviews empirical evidence of their effectiveness. Furthermore, this article aims to identify the problem at the heart of IEWV, putting it in context of the literature on strategic games (section \ref{sec:Principal-Agent problem}), which establishes common ground and an intimate connection among mechanisms.
%|% How this study is different from other studies: There are a couple of reviews, but none of them is as comprehensive nor do they discuss relations between mechanisms.
Two other works also review this literature \parencite{charness_experimental_2021, faltings_game-theoretic_2023}, but both review only a few mechanisms for IEWV.

%|% What this study found: Mechanisms are not ready to be applied off-the-shelf
This study finds that implementing these mechanisms in science, policy or business cannot be recommended \textit{yet}, 
%|% Reason: 
because the empirical evidence regarding the the mechanisms' effectiveness is sparse and weak.
%|% %Claim: A major cause of this is mechanism complexity. 
The primary cause of this is mechanism complexity.
%|% Reason: Mechanism designers have tried to devise ever more clever designs, at the cost of being able to communicate them to research subjects.  
If respondents do not understand how their reports translate to their rewards, the incentives are ineffective. Most mechanisms are considered \say{extraordinarily difficult to fathom} \parencite{charness_experimental_2021} and cannot be easily conveyed to respondents.  Consequently, mechanism complexity inhibits both research and application.

% address shortcomings of the current literature and its causes.  
% clearly outline future research directions \ref{sec: Discussion}
In discussing the current research, this article also suggests many directions for future research. As mechanism complexity impedes empirical analysis, a major future research direction should be the creation and assessment of easier-to-comprehend mechanisms. Another major research direction is the large-scale empirical validation of these mechanisms to provide convincing evidence on whether, and why, mechanism do or do not improve the accuracy of reports. 

%|% Where to read what in this article
The rest of this article is organized as follows: Section \ref{sec:Principal-Agent problem} defines the problem. Section \ref{sec: Mechanisms} reviews some of the most widely discussed mechanisms and existing empirical evidence of their effectiveness. The tables \ref{tab:Truth_serums}, \ref{tab:Market_based} and \ref{tab:Output_Agreement} list all reviewed mechanisms and can be found in the appendix. Section \ref{sec: Applied research} gives a brief overview of some research projects in which mechanisms were used to create incentives for truth-telling. Section \ref{sec: forecast_combination} reviews articles that use these mechanisms for forecast combination, a related problem.  Section \ref{sec: Discussion} critically discusses the existing research and proposes concrete future research directions. Section \ref{sec: Conclusion} concludes.

%-----%

\section{Belief elicitation as a Principal-Agent problem}\label{sec:Principal-Agent problem}

%This section gives the reader a deeper understanding of what the actual problem is

%|% Context: %|% Concrete modeling of the situation 
%|% Assume a principal (principal) that wants to \textit{purchase} information. 
We assume a principal who would like to gather information
%|% The principal has no access (variation: costly) to the information. 
to which he lacks access,\footnote{The principal has a male identity and the agents a female identity throughout the paper, as is common in the literature on non-cooperative games.} e.g. a journal editor seeking to evaluate a paper for publication.
%|% An agent or expert (or consultant or seller) offers to sell the information that the principal seeks - The expert has noisy access to the information 
Multiple other agents (e.g. peer-reviewers), henceforth called 'respondents', receive a signal of the information and may report this to the principal.\footnote{There are many different names in the literature for the roles of principal and agent. The principal is often also called the center or receiver. The agent is often called the subject, seller, consultant, player, worker, expert, forecaster, rater, or sender.}

%|% Claim:
%|% MAIN CLAIM : Although the principal is unable to verify reports, the principal can make contracts that incentivize the expert to be truthful. This is possible only when there are multiple experts that have access to the same information (although different beliefs can result from noise). The principal can purchase multiple reports and compare them against each other, the price is determined depending on other reports. 

\textbf{Although the principal is generally unable to verify the accuracy of any single report, the principal can create contracts that \textit{incentivize} the respondents to truthfully share their beliefs.}
%|% Reason: 
This is possible because the principal can compare reports with other respondents' reports when multiple respondents have access to the same information. As we will see in section \ref{sec: Peer-Prediction and BTS}, this holds true even when these respondents receive only noisy signals of the information and thus will hold different beliefs.

%|% Claim: IEWV mechanisms should not be used when verification is possible.

We call incentive structures that try to achieve this \textit{IEWV mechanisms}. These mechanisms should not be employed when verification of the truth is possible,
%|% Reason: 
because the rewards to the respondent depend only on other respondents' reports. Consequently, respondents are playing a \textit{game} with one another.\footnote{\textcite{ottaviani_strategy_2006} put this the following way: \say{Clearly, there is no gain from conditioning the state-contingent reward on the messages of
other forecasters when the realized state is sufficient for such messages. When instead the
state is observed with noise, or equivalently when forecasters possess conditionally dependent signals, conditioning the reward also on the competitors’ forecasts might improve the
incentives for forecast accuracy}.}
%|% Evidence:
We will see in section \ref{sec: Peer-Prediction and BTS} that respondents can collude to manipulate their rewards \parencite{gao_trick_2014}.
%|%Claim
Whenever truth is available, mechanisms that compare reports directly to truth, such as e.g. bets or proper scoring rules, are superior, %|% Reason: 
as stating the true belief is a dominant strategy and does not depend on others' reports \parencite{parmigiani_decision_2009}.\footnote{Truthfulness is  a dominant strategy under strictly proper scoring only if respondents are risk-neutral utility maximizers \parencite{gneiting_strictly_2007}. A growing literature concerns itself with relaxing the assumption of risk-neutrality \parencite{offerman_truth_2009, hossain_binarized_2013}.}

%|% Thus, being truthful becomes a \textit{game}. This is different when reports are directly compared with true outcomes, only the outcome matters for the individual report. Being truthful is a \textit{dominant strategy}, if proper scoring rules are deployed. \textit{IEWV mechanisms should not be used when verification is possible.}
 %|% Since the payoff is only determined by others reports, the corresponding state in which everyone is truthful is the desired Bayesian Nash Equilibrium (BNE). 

\begin{figure}[h]
    \centering
    \includegraphics{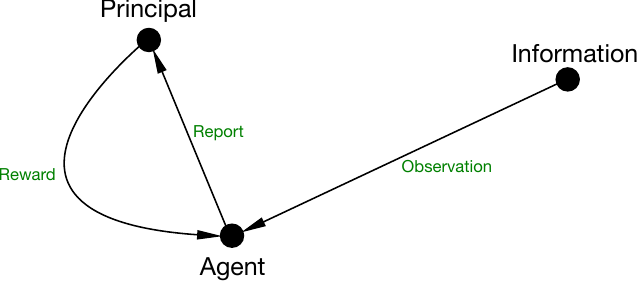}% Change to INFORMATION OF INTEREST
    %Change to principal (Principal) and SUBJECT (Agent)
    \caption{Information elicitation without verification (IEWV) is a principal-agent problem}
    \label{fig:Principal_Agent_problem}
\end{figure} 

%|% We find ourselves faced with the classical principal-agent problem of \textit{hidden information}. The principal does not know how valuable the information that the expert offers is and thus they will have difficulty of converging on a price for that information.
This scenario exemplifies a principal-agent problem with information asymmetry, specifically \textit{hidden information}: The principal does not know how valuable the information that the respondent offers is, because he is uncertain regarding the respondents truthfulness.

%|% However, the expert does not care about the decision at all and gains utility only from her payoff.
The respondent gains utility only from the reward that the principal offers, and the respondent does not care what the principal does with the information provided.\footnote{%|% IEWV is highly related to games of strategic information transmission - where the agent/subject has a stake in the decision. For example, a worker reporting to their boss may have an interest in the bosses decision. In the situation of IEWV, the subjects do not care about any decision made as a result of their report, but they only report truthfully and effortful if they have an incentive to do so. If the subjects also cared about the outcome of the decision, this would yield wildly different behavior. CITE Osbornes game theory
The IEWV problem is highly related to games of strategic information transmission. However, in games of strategic information transmission, the respondent also cares about the final decision. For example, an advisor to a policy-maker usually has political beliefs of their own and may try to influence final decisions for what they believe to be better. Thus, there is an additional incentive to provide (non-)truthful answers \parencite{osborne_introduction_2004}. Therefore, games with strategic information transmission yield very different behavior and different mechanisms are needed to align incentives.} The directed graph in Figure \ref{fig:Principal_Agent_problem} describes the interaction.
%|% The expert can choose to submit her informed beliefs to the principal. If the expert does so, she is "truthful" (that is her type). (There is no way of credibly signaling the type as it can be switched at no cost)
Any respondent can choose to submit the information to the principal. If the respondent does so, she is being \textit{truthful}. She can at no direct cost to herself also submit any other information and be \textit{non-truthful}. There is no way for the respondent to credibly signal her type.

%|% The principal knows what the value of perfect unbiased information is, but cannot verify whether information is accurate or truthful. He does not know the experts type. Thus, he cannot pay the expert based on the accuracy of the reports. 
%List those cases here with explanations and a short example each
There are exactly 4 classes of IEWV problems that may be distinguished based on the information structure of the problem. This is because there are three directed arrows in the graph that describes the game, which makes for $2^3=8$ possible combinations, but only four of them describe the situation that we are interested in.\footnote{Specifically, the following four cases are irrelevant: (i) There is no information, (ii) the principal has access to information, (iii) the agent has access to information but does not share it and (iv) the principal has access to information and the agent shares her (non-informed) belief.}
 All of the four relevant combinations have in common that the principal cannot compare the respondents report against the true outcome immediately.

\begin{figure}[h]
    \centering
    \includegraphics[width=.9\textwidth]{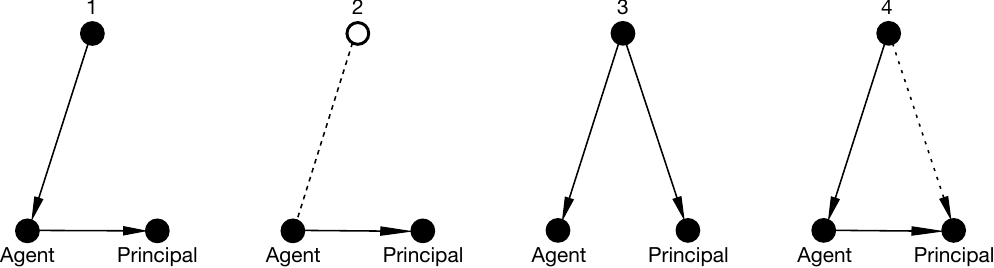}  
    \caption{Multiple cases in which verification and evaluation of correct signaling cannot occur. Dashed lines indicate temporal delays.}
    \label{fig: Different cases of unverifiability}
\end{figure}

\begin{enumerate}
    \item \textbf{The information is only ever observable by the respondent.} This situation is depicted in point 1 in Figure \ref{fig: Different cases of unverifiability}. The arrows signal the direction in which information is transmitted. 
    
    Example 1 | Self-reported data: A drug trial is conducted and the respondents must report which side effects they experienced, or whether they consumed alcohol during the trial period.\footnote{Side effects of drug usage are essentially a noisy signal of the drug's side effects in the population. Thus, IEWV mechanisms can incentivize truthful answers to such self-report questions. Truthfulness can be incentivized in the case of individual-specific alcohol consumption as the answer can be used to condition on alcohol consumption. Then, correctly stating alcohol consumption and corresponding side effects should better predict the side effects of others with the same alcohol consumption, and thus increase the respondent's own reward.}

    Example 2 | Causal effects:
    Policy decisions are often informed by expert opinion, with policy-makers having less insight into the likely consequences of their decision than the experts. Policy-makers need to trust the experts. They are unlikely to get evidence on the consequences of the decision not taken and thus cannot compare the information provided by the expert to actual outcomes.

    Example 3 | Long-term forecasts: Although long-term outcomes are observable, the revelation will only occur in such a distant future that we may not live to see it.

    \item \textbf{The information is currently observable by the respondent and will be observable by the principal.} This situation is depicted in point 4 in Figure \ref{fig: Different cases of unverifiability}.
    
    Example | Short-term forecasts: A managerial decision-maker in the firm would like to forecast demand for a product in the next quarter. This information will be unveiled after the quarter. The decision-maker eventually has access to the information but will make use of the temporal precedence of the respondents beliefs as a proxy for the true demand to make decisions now. Elicitation thus takes place without immediate verification, but verification will occur at a later time. As discussed previously, mechanisms such as Proper Scoring Rules allow the principal to reward accurate reports \parencite{gneiting_strictly_2007}.\footnote{This is still true if verification is \textit{unlikely}, as proper scoring rules retain their properties when multiplied with a constant (the probability of evaluation) \parencite{gneiting_strictly_2007}. That is, if there is a non-zero chance that a report will be scored properly upon evaluation, truthfulness is a dominant strategy. A principal could incentivize truthful and effortful reports of information that he can only gain with great effort or expense. For example, beliefs regarding the replicability of 10 studies could be elicited truthfully as long as there is a chance that at least one of them will be replicated.}
    Forecasting can thus be viewed as a special case of IEWV.\footnote{Although much more literature is concerned with forecasting than IEWV in general, forecasting remains a special case in the sense that it is actually the exception rather than the rule that short-term verification is possible \parencite{faltings_game_2017}.}
    
    \item \textbf{The information is no longer observable, but the respondent did observe it in the past.} This situation is depicted in point 2 in Figure \ref{fig: Different cases of unverifiability}.

    Example | Reconstructing the scene of a crime: A police officer is tasked with reconstructing the scene of a crime. Given that he was not present at the scene itself and that the crime has already occurred, he must rely on reports from eyewitnesses.
    
    \item \textbf{Future observation of the truth by the principal will render rewards meaningless.} In this case the situation would change from the point 1 in Figure \ref{fig: Different cases of unverifiability} to point 3, where evaluation and rewards cannot be transferred, if the event occurs.

    Example 1 | Risk of collapse: A managerial decision-maker would like to know the probability that the firm will go bankrupt next quarter. The respondent cannot expect to receive any pay if the firm goes bankrupt. Thus, a respondent that is only paid after the end of the quarter has a strong incentive to report zero probability of bankruptcy, regardless of the true risk. The respondent's incentive is to maximize the accuracy of her report in the only scenario she cares about: when the firm is not bankrupt.

    Example 2 | Global catastrophic risk: Similarly to the previous example, if a policy-maker is interested in eliciting the probability that a pandemic will cause a major catastrophe, experts will find the prospect of post-catastrophe rewards not engaging.

\end{enumerate}

%Incorporate the case 'Forecasting' by adding A* slightly later than A and drawing an arrow to D
% Or simply cut out time entirely, as it confuses anyhow. Just draw the four cases!

%|% The principal can purchase the information from the expert unconditionally and will do so if he expects it to be accurate (with high enough probability). This is called "introspection".
The principal can purchase the information from the respondent unconditionally and will do so if he expects it to be sufficiently accurate. This approach is called \textit{introspection} - i.e. simply asking for information without incentivizing truthfulness.
%|% Why would the expert ever lie?
%|% (variation: costly) The expert can choose to engage in additional effort to obtain more information and submit a strictly more accurate belief. 
One might ask why the respondent would ever be untruthful. However, anyone who has ever tried to accurately report their own belief will find that this requires non-zero effort. We could extend the model to incorporate effort. 
%|% It is effortful for the expert to obtain (additional) information
Then, the respondent would only obtain (additional) information at a cost $c$. If the respondent obtains the information at this cost, she is being \textit{effortful}. It would be irrational to be effortful (and truthful for that matter) for any $c>0$, because the respondent is not rewarded for the additional effort.\footnote{%|% The principal can obtain the information himself, at a greater cost. 
Further, one could extend the model to assume that the principal can also observe the information at a cost $c_d$ , which is strictly greater than the cost of obtaining the information is to the respondent. An example of this approach is provided by  \textcite{baillon_peer_2025}.}

%---------------------------------------2 THEORETICAL RESEARCH---------------------------------------------------------------%
\section{Mechanisms}\label{sec: Mechanisms}

%%----------------------- Output Agreement ----------------------%%
\subsection{Output Agreement \& Proxy Scoring}

\subsubsection{Output Agreement}

%|%Output Agreement is the simplest possible game to incentivize truth-telling
Rewarding the respondent if her report is identical to other reports is the simplest approach to reward unverifiable reports.
%|% Output agreement has been applied in practice
\textcite{von_ahn_labeling_2004} are the first to both propose and apply such a mechanism. The authors create a computer game where two players are both shown an image and are asked to label it. The players are rewarded if the descriptions of the image match. To avoid players colluding, some words or single characters are taboo and players are matched randomly for each image to avoid repeated interaction.
The game has sparked wider interest in human computation via games \parencite{von_ahn_designing_2008, law_human_2011, huang_enhancing_2013}. Particularly, the game has addressed a major issue in artificial intelligence: how to inexpensively label large amounts of hard-to-verify data. This application remains highly relevant. 

%|% Empirical analysis ?

%|% Output Agreement is promising and simple, but has major fundamental drawbacks 

These mechanisms have the benefit of being easy to describe. However, \textit{Output Agreement} schemes that directly match answers do not incentivize respondents to share private information \parencite{waggoner_output_2014}.\footnote{Output Agreement must not be confused with Outcome Matching - a different elicitation method for when the truth is verifiable \parencite{charness_experimental_2021}.} A respondent who possesses private information will not choose to reveal it, as it is unlikely to be featured in another respondent's report. For example, if the image to be labeled shows a lynx, and the respondent knows that this is an image of an iberian lynx, but finds it unlikely that others will recognize this, they are better off just reporting \say{lynx}. As a result, Output Agreement games are actually disguised guessing games, in which rewards are maximized by guessing majority opinion. In fact, common knowledge is the best result that can \textit{theoretically} be hoped for \parencite{waggoner_output_2014}, unless one assumes respondents to exhibit a bias where they falsely assume their own belief to be the majority opinion, known as the \say{false-consensus effect} \parencite{carvalho_inducing_2017}.  

\subsubsection{Proper Proxy Scoring Rules}

%|%Proxy scoring rules - also has the wisdom of crowds benefit

%|% Claim

\textit{Proper Proxy Scoring Rules} improve upon Output Agreement by strengthening the incentives for truthfulness: Instead of rewarding agreement, respondents are rewarded for reporting a \textit{combination} of others' reports \parencite{witkowski_proper_2017}, utilizing the wisdom-of-crowds. For example, respondents could be properly scored against the mean of all estimates.
%|% Warrant:  Wisdom-of-crowd effect - What is it.
The wisdom-of-crowds is a phenomenon that occurs when respondents' reports are combined to form a single consensus estimate. Such a consensus estimate is usually better (in expectation) than any individual report from within that group of respondents
\parencite{surowiecki_wisdom_2004, clemen_combining_1989}. 
%|% Reason: 
Proper Proxy Scoring Rules may be effective because if all respondents perceive the combined measure as a \textit{better estimate} of truth than their own belief, the respondents' reward is maximized for reporting their belief if everyone is truthful. The combined estimate acts as a proxy for the truth.
%|% Acknowledgement&Response:
\textcite{witkowski_proper_2017} acknowledge the fact that this mechanism is not fully incentive-compatible in theory, and primarily motivate it in the context of standard forecasting.

%|% The wisdom-of-crowd effect leads to the Nash Equilibrium

%|% The paper on the fusion mechanism also proposes to score against combined outcomes. 
\textcite{papakonstantinou_mechanism_2011} propose a similar mechanism in which reports are combined to obtain an aggregate measure that each individual report is scored against. The paper assumes that respondents report Gaussian distributions over a real variable and that respondents need to engage in costly effort to obtain signals. By conducting a second-price auction prior to elicitation of reports, the payouts are scaled such that the cost of effort is guaranteed to be compensated. The Gaussian distributions are added up such that the combined measure that reports are compared against resembles the mean. The authors claim that this provides incentive-compatibility.

\subsubsection{Empirical evidence on Proper Proxy Scoring rules}

\textcite{atanasov_improving_2025} study how different scoring rules affect the respondents' judgment regarding low-probability events.\footnote{Since low-probability-events are by definition rarely ever observable, IEWV mechanisms can be an effective way of incentivizing honest responses.} They find that Proper Proxy Scoring rules significantly improve the accuracy of reports compared to introspection, on par with the squared error/Brier score.
% two information aggregation mechanisms for predicting ....
\textcite{court_two_2018} study beliefs regarding box office revenues of Australian movies.\footnote{The authors call this mechanism \say{Guess of Guesses}, and do not cite \textcite{witkowski_proper_2017}. That the method coincides with Proper Proxy Scoring Rules seems to be unintentional.}  Respondents received a higher reward if their reports were closer to the median report. However, it is not quite clear from the paper to what extent study participants were made aware of this. The study finds that the respondents predicted box office revenues better than a random guesser. However, when the same respondents answered different questions and their estimates were directly compared with true outcomes, their reports were much more accurate. It is not clear whether this should be attributed to the mechanisms or the difference in difficulty between the two sets of questions. The study on reciprocal scoring (discussed in the following section) can also be viewed as evidence for the effectiveness of Proper Proxy Scoring rules, given their similarities.

\subsubsection{Reciprocal Scoring}

%Reciprocal Scoring is Output Agreement in groups
\textit{Reciprocal Scoring} is a variation of Proper Proxy Scoring \parencite{karger_reciprocal_2021}. Respondents are randomized into groups. The respondents are then asked to provide their reports. After all respondents have submitted their reports on the question, the groups' median report is computed.\footnote{Reciprocal scoring may very well work with other combinations of estimates such as e.g. the mean.} Respondents within a group are rewarded based on how close the median report is to a reference groups median report.
%|% But there is a wisdom-of-crowd effect too
Therefore, Reciprocal Scoring is Output Agreement in groups. However, the median report may display a wisdom-of-crowd effect too. Thus, there is a stronger incentive to report private information, as with Proper Proxy Scoring.
%|% Colluding is more difficult in groups
Moreover, colluding is arguably more difficult in groups.

%|% Empirical analysis
\subsubsection{Empirical evidence on Reciprocal Scoring}

%Reciprocal Scoring
%|%Claim:
\textcite{karger_reciprocal_2021} also tested their proposed Reciprocal Scoring in two separate empirical studies, with very promising results. Reciprocal Scoring causes reports to be significantly more accurate compared to introspection. The first study serves to investigate how forecasters respond to the incentives posed by Reciprocal Scoring. The second study showcases the use of Reciprocal Scoring by providing estimates on the causal effect of different policy measures on COVID-19 deaths. In the first study, the authors conducted a randomized trial. 1284 respondents recruited via Prolific participated in ten forecasting tasks. The respondents were randomly assigned to three different mechanisms: Introspection, Proper Scoring, and Reciprocal Scoring. Although the authors did not confront respondents with the exact workings of their mechanisms, they tried to convey the general intuition of the different mechanisms by using examples. The respondents assigned to Reciprocal Scoring were informed that they will maximize their earnings if they accurately predict the predictions of a set of separately recruited proficient forecasters, called \say{superforecasters}.\footnote{The term \say{superforecaster} has been introduced by Philipp Tetlock's research on psychological traits of accurate forecasters \parencite{tetlock_superforecasting_2015}. Superforecasters were scored properly against true outcomes, but the participants were not told how superforecasters were scored.}

\begin{figure}
    \centering
    \includegraphics{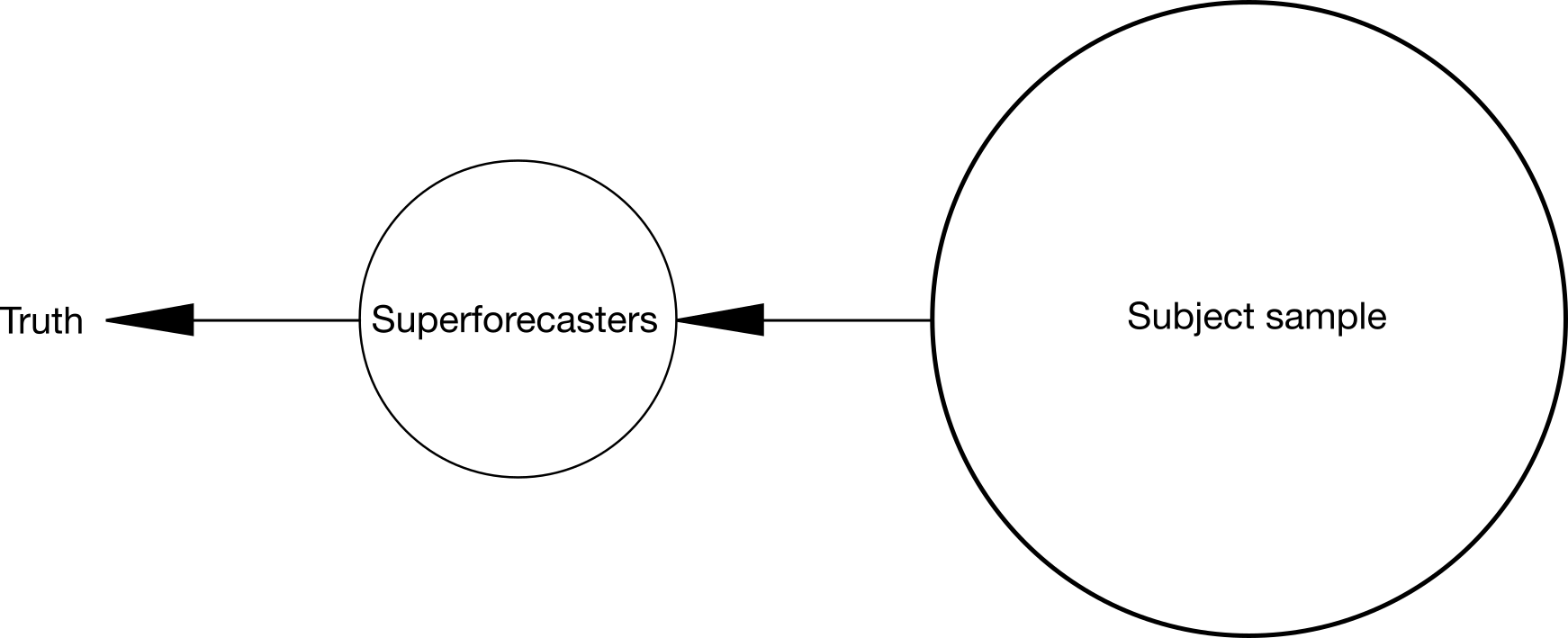}
    \caption{How respondents were scored in the study by \textcite{karger_reciprocal_2021}}
    \label{fig:reciprocal_scoring_implement}
\end{figure}

 Instead of splitting the respondents into two groups that would predict each other, the authors implemented a separate superforecaster team as the reference to be predicted. This experimental design has the merit of retaining full sample size. However, it is different from the core idea of dividing the sample up into at least two equal groups. What the study actually measures is whether respondents recruited via Prolific behave differently when evaluated using a scoring as described by Figure \ref{fig:reciprocal_scoring_implement} compared to proper scoring  or introspection. Respondents might have assumed that superforecasters' predictions proxy truth closely, and much better than their own belief. If so, they had a stronger incentive to be truthful in this setup than if they had to predict another group with similar predictive skill. In this sense, the experimental setup is much closer to Proper Proxy Scoring Rules than to Reciprocal Scoring itself.

 Reciprocal Scoring leads to slightly less accurate reports than proper scoring, but this difference is insignificant. Respondents that were scored with Reciprocal Scoring took as much time to answer questions and consulted as many sources as the respondents that were scored with proper scoring. It would be most interesting to see whether Reciprocal Scoring is equally effective outside of the study's specific setup. Skilled forecasters may respond differently to this situation and may be more adept at colluding. Furthermore, it is not clear whether experts will be discouraged from sharing private information in this setup, as it may be unlikely that this information is reflected in the reference group's median report.

In the second study by \textcite{karger_reciprocal_2021}, two teams of experienced forecasters are asked to predict the other teams' forecasts regarding the effect of COVID-19 policy measures on COVID-19 deaths. The teams' median reports are relatively similar and show large differences in effect size across policy interventions. Since the outcome is not observable, this study simply demonstrates how Reciprocal Scoring would be implemented in practice.

%---------------------------------------2.2 Peer-Prediction ---------------------------------------------------------------%
\subsection{Peer-Prediction Mechanisms and Truth Serums}\label{sec: Peer-Prediction and BTS}

\subsubsection{Peer-Prediction}

%|% Main claim: 

\textit{Peer-Prediction} tries to improve upon Output Agreement by incorporating the \textit{belief formation process} into the scoring \parencite{miller_eliciting_2005}.  This has its merits,
%|% Reason:
as \textcite{zhang_elicitability_2014} show that \textit{some} assumptions regarding the belief formation process need to be made in order to arrive at the result that truth-telling is a strict Bayesian Nash Equilibrium (BNE) of a game, i.e. the mechanism is incentive-compatible. 
However, explicitly modelling beliefs leads to practical downsides as it introduces additional assumptions and complexity.%|%Claim + Reason : 

%|% Evidence: Peer Prediction assumes a concrete information structure 
Peer-Prediction assumes a most simplistic situation: It is assumed that all respondents have a common prior belief and observe a \textit{noisy} signal of the information of interest. This situation shall be called the \textit{common-prior-single-signal} (CPSS) case. Most mechanisms reviewed throughout this text make similar assumptions. The reported beliefs (if truthful) should be strongly correlated because the signals that respondents received have a common cause. 

For example, consider an academic peer review. If the submitted paper is good, their ought to be many positive peer reports. If the paper is bad, more reports should be negative.\footnote{The authors call this 'stochastic relevance'. This means that the beliefs are strictly correlated with the information. Different information leads necessarily to different beliefs.} Although the principal (here the journal editor) does not observe the actual truth (paper quality), they can make use of the chain $\textit{Information of interest}\rightarrow \textit{respondents beliefs} \rightarrow \textit{reported beliefs}$ to gather the unobserved information.
A full numerical example can be found in the Appendix.
Essentially, the editor (principal) is asking: What is the probability that another (reference) reviewer will recommend to publish the draft? The answer to this question is evaluated using proper scoring. Assuming a common prior belief that 20\% of reviewed papers are good, a reviewer that liked the paper would report a probability greater 20\%. A reviewer that did not like the paper would report a probability smaller than 20\%. 

 %|% Essentially, Peer-Prediction improves upon Output Agreement by rewarding agreement on probabilities.
 Peer-Prediction rewards agreement on probabilities.
 %|% Peer-Prediction does not require that subjects explicitly estimate probabilities
 But Peer-Prediction does not require that respondents explicitly estimate probabilities.\footnote{Peer-Prediction works for non-binary information too. It is not required that there are only two (or n) states.} By simply asking whether the reviewer liked the paper, and moving the Bayesian updating to the scoring function, reviewers can simply report their belief (reject/accept), thereby implicitly making the probabilistic Peer-Prediction. This greatly simplifies the practical application. However, this requires the principal to know the prior \textit{and} how respondents are updating given their private information.\footnote{That is, the posterior belief is identical across respondents who have received the same signal. This assumption is called impersonal updating.} These assumptions are unlikely to hold in most real-world applications.
%|% Matching probabilities instead of choices can be gamed equally well. 
A substantial body of literature points out that matching probabilities can be gamed just as Output Agreement can, and the respondents have an incentive to collude as this yields higher rewards \parencite{gao_trick_2014, dasgupta_crowdsourced_2013, shnayder_informed_2016, kong_equilibrium_2018, faltings_game_2017, jurca_mechanisms_2009}. For example, if all reviewers simply decline all drafts, they would receive at least as high a reward as if they reported truthfully. The truthful BNE is not necessarily the one with the highest expected utility.

%|% Empirical analysis
\subsubsection{Empirical evidence on Peer-Prediction}

% Peer-Prediction 

\textcite{gao_trick_2014} are the first to put Peer-Prediction to the test. Respondents recruited via MTurk play a minimalistic Peer-Prediction game. A key difference between the game's setup and the original design is that the respondents play the game \textit{repeatedly}. Furthermore, the respondents only receive binary signals and submit binary reports, so the game takes the simplest possible form. The researchers do not explain the mathematics or the intention behind Peer-Prediction but they do explain the general mechanism and explicitly calculate and display potential earnings to respondents. The latter is possible because the game is so simple. There are only four potential outcomes per round of play. The result of the study is that respondents quickly begin to coordinate their reports around non-truthful but higher-paying equilibria. In other words, the respondents quickly begin to game the system. The researchers compare the behavior to a control group that is rewarded with a fixed pay (introspection) each round. They find that introspection yields strictly more truthful reports than the Peer-Prediction mechanism. In this sense, Peer-Prediction makes things worse by incentivizing coordination on non-truthful equilibria, whereas respondents who have no incentive to be truthful often choose to do so. However, this finding comes with a couple of caveats. The game is set up so that it is very straightforward for respondents to game the system. Respondents can learn to coordinate over many rounds, which they do. In practice, this could be avoided. Secondly, there is no effort involved in truth-telling. The respondents simply need to report the signal that they have received. This heavily favors introspection in this analysis. In real-world applications that do not only involve honest rating, but e.g. forecasting complex events, truthfulness often comes at a significant cost.\footnote{In the case of forecasting complex events, this cost would be the time and energy spent researching the subject matter.} Since introspection does not incentivize investment in obtaining additional information introspection may perform relatively less well on more complex tasks. 

\textcite{mandal_effectiveness_2020} attempt to study the effectiveness of Peer-Prediction mechanisms for long-term forecasting, but fail to provide convincing evidence. The study is a most honorable quest, as long-term forecasting is one of the main underexplored IEWV problems \parencite{gruetzemacher_forecasting_2021}. Unfortunately, the study's methodology is not related to long-term forecasting at all, except for the dubious conjecture that \say{since the hybrid scheme also improves user engagement, this suggests that the hybrid scheme would provide the best accuracy for longer term forecasting events}.\footnote{The hybrid scheme refers to one of the treatments where rewards depend on a mixture of proper scoring and a version of the Peer-Prediction mechanisms called Correlated Agreement (see section \ref{sec: Variations of BTS + PP})} 
Furthermore, the average prediction error in \textit{all groups} of the randomized trial with respondents recruited via MTurk is worse than that of a random guesser. The respondents possessed no foresight. Lastly, the authors did not match respondents directly but compared reports with averaged values of reports sourced from a forecasting tournament \parencite{ungar_good_2012}. This sabotages the whole idea of the study, as respondents from the forecasting tournament are incentivized with proper scoring. In other words, the respondents are not actually playing against each other, but against other respondents from the tournament, who are incentivized to be truthful. Furthermore, it is not clear how respondents were instructed, or whether they did comprehend the mechanisms.

%---------------------------------------2.1 BAYESIAN TRUTH SERUM---------------------------------------------------------------%
\subsubsection{Bayesian Truth Serum}\label{sec: BTS}

%|% Claim:

The Bayesian Truth Serum, independently proposed by \textcite{prelec_bayesian_2004}, is very similar to Peer-Prediction,
% follows a similar idea, but instead of requiring the principal to know the common prior up ahead, the BTS elicits it
% How the BTS differs from the Peer-Prediction Method
but it does not require that the principal to know the common prior. Instead, the principal learns about the common prior from the respondents. This is beneficial, but introduces additional complexity. The respondents are each asked two questions: 

\begin{enumerate}
    \item What is your belief?
    \item What is your prediction on the distribution of others' answers (on question 1)?
\end{enumerate}

Notice that the second question \textit{is} the Peer-Prediction. We return to the example of academic peer review: The reviewer would state whether they recommend publishing the draft (Yes/No) \textit{and} what the probability that others do so is. The Peer-Prediction mechanism is able to elicit the information in one question because the common prior is assumed known.
With the BTS, a respondents reported private belief and their Peer-Prediction together \textit{imply} the common prior \parencite{witkowski_peer_2012}.\footnote{If one were to take the common priors assumption seriously, it would actually be sufficient to ask the second question to just one of the respondents.} Thus, the prior need not be known to the principal.

%|% Claim: 
A major drawback of the BTS is that it requires a large pool of respondents
%|% Reason: 
because the Peer-Predictions (second question) are compared against the actually observed \textit{frequency} of reported beliefs (first question). 
%|% Example:
For example, a reviewer may decline a draft and report that 1 in 16 reviewers will recommend publishing it. The actual frequency of publish/decline reports is compared to the report, requiring at least 16 responses.
%|% The BTS requires many participants to calculate frequencies
Peer-Prediction has the major advantage over the BTS that it works with as few as three respondents, because the reports are not compared to the entire group's reports, but to randomly matched respondents. 
%|% Acknowledgement&Response: 
However, the BTS can be amended to work with as little as two respondents \parencite{cvitanic_honest_2024, radanovic_robust_2013, witkowski_robust_2012}.

%|% The BTS pays based on predicting surprisingly commonly held beliefs - which, given some strict and simplified additional assumptions, is better than pure output agreement for incentivizing truth-telling

%|% CLaim: 
The BTS is exclusively applicable to questions that have categorical answers 
%%% Reason: 
as the response to the initial question is determinate rather than probabilistic.
%|% Common priors will be rare to find for estimation tasks 
The BTS has the big advantage over Peer-Prediction that it allows to easily check whether priors are indeed the same, but no study has reported doing so.
Publicly available datasets collected for studies such as \textcite{palley_boosting_2023} show that respondents often have very different priors regarding the variable in question.
As with Peer-Prediction, other researchers have pointed out that the BTS has non-truthful equilibria that pay at least as much as honest reporting, i.e. respondents can collude to achieve higher rewards \parencite{jurca_mechanisms_2009, waggoner_output_2014}.

\subsubsection{The intimidation method}\label{sec: intimidation method}

%|% Mechanism complexity becomes a real issue
The BTS has been criticized for its unrealistic assumptions and high complexity \parencite{charness_experimental_2021}. 
%|% The suggested solution - "just be truthful"
The author acknowledges the inherent complexity of the mechanism in the paper and suggests that the BTS must not be explained to respondents, but that respondents can instead be reassured that they maximize their rewards by being truthful.\footnote{This is similar to the use of proper scoring in forecasting. Proper Scoring rules are complex too, but respondents rarely need to calculate payouts because more accurate responses yield strictly higher expected value. Being truthful is a dominant strategy.}
%|% Being truthful is a Bayesian Nash Equilibrium for the BTS

%|% Claim: 
For IEWV, however, claiming that rewards are maximized by truthful reports is potentially untrue, and thus deceptive.
%%% Reason: 
Rewards are maximized in expectation only if \textit{everyone} reports truthfully \textit{and} all assumptions hold. This can clearly never be guaranteed to respondents ex ante.\footnote{It can be checked ex post:  In at least one study \parencite{zhou_long-term_2019} respondents did not maximize their rewards in expectation if they were truthful.}
%%% Evidence: for it does not matter which actual payment scheme is implemented. 
It does not matter whether a mechanism is implemented at all, as the method relies on the respondents blindly trusting the principal's claim \parencite{charness_experimental_2021, schoenegger_taking_2022}.

%|% The intimidation method is deceptive to some degree 
Informing respondents that they maximize their rewards by being truthful and leaving them in the dark about the actual mechanism is henceforth called the \textit{intimidation method}.\footnote{%|% There are certain similarities to the bogus pipeline which is also based on deception, and worked well
 The intimidation method is similar to the \textit{bogus pipeline}, a technique in which respondents are told that they are hooked up to a lie-detector, and that has been successfully used in psychology research for decades \parencite{roese_twenty_1993}.} 
%The intimidation method is simple to implement
Clearly, the intimidation method is the simplest way of eliciting beliefs, if respondents believe the claim. 
%|% The intimidation method may lead subjects to mistrust researchers claims, if used excessively. This would be quite bad. 
If used excessively, respondents may learn that the claim is not fully correct and start to mistrust the principals instructions, which would be a very undesirable consequence.
%|% Ethical considerations are not yet present, but needed. 
Ethical considerations regarding the use of such methods are lacking, but needed.

%|% Many have rushed to empirically validate the BTS... by using the intimidation method. This only provides us with evidence regarding the intimidation method, as the subjects have had no idea of the BTS. Ironically, the intimidation method is well tested empirically, much better than any of the other mechanisms reviewed in this article.
A handful of studies have attempted to empirically validate the BTS. Since these studies actually just tell respondents that they will maximize their rewards when they are being truthful, this only provides us with evidence regarding the intimidation method. Ironically, the intimidation method is much better empirically tested than the BTS or any other mechanism reviewed in this article. The bottom line is that the intimidation method is effective and robustly improves the accuracy of reports.

\begin{figure}
    \centering
    \includegraphics[width=0.8\textwidth]{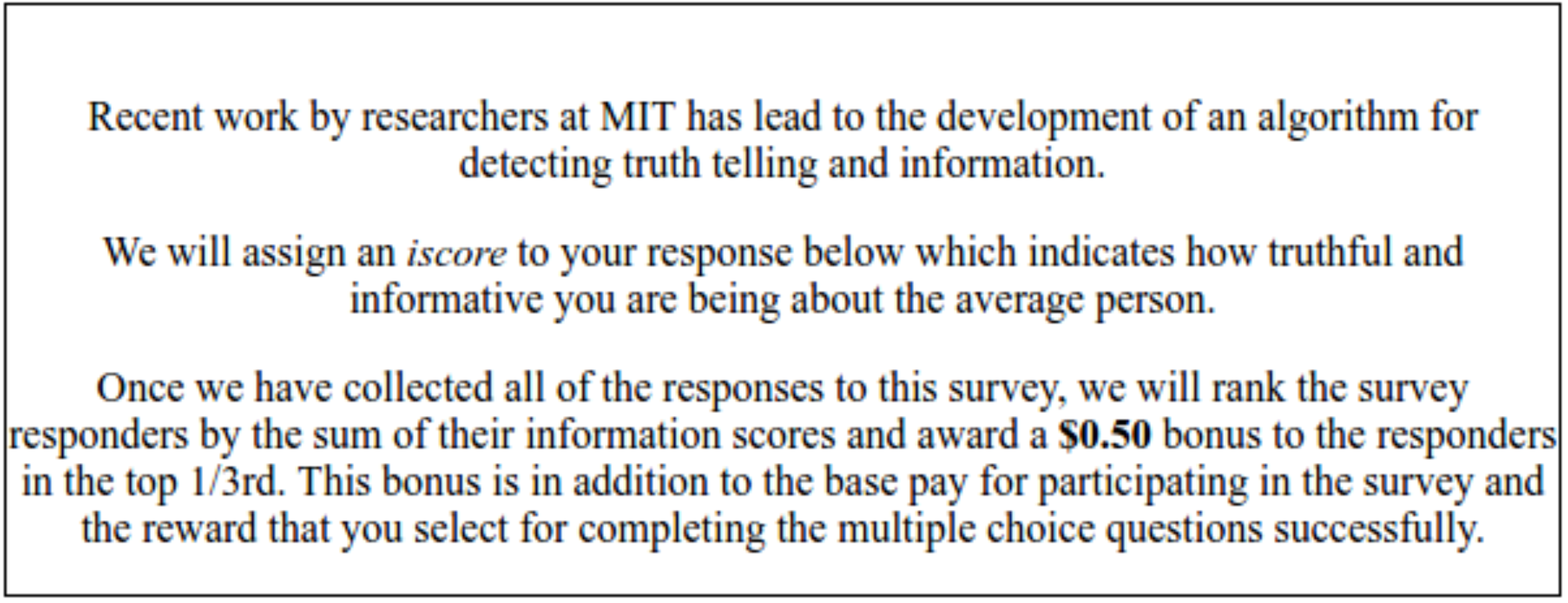}
    \caption{Information displayed to respondents in \textcite{frank_validating_2017}}
    \label{fig:Intimidation}
\end{figure}

\textcite{frank_validating_2017} try to validate the truth-telling incentives of the BTS in large-scale online experiments. Their statement to respondents is detailed in Figure \ref{fig:Intimidation}. The authors run a randomized trial with one group being subject to the intimidation method, one group that in addition to the intimidation treatment receives dynamic score feedback after each report and one control group. \textcite{frank_validating_2017} ask respondents to flip coins and roll dice and report their results. A bias is induced by paying extra for heads and higher integers when throwing dice. The main result is that the intimidation method succeeds in lowering the induced bias. The dynamic score feedback has not much of an effect; most of the boost in honesty comes from the intimidation. The methodology is not well-suited to test any IEWV mechanism because the outcomes of dice rolls and coin flips are public knowledge. There is no reason for respondents to actually flip a coin, i.e. obtain a signal of the information of interest. However, it fits nicely into the literature on preferences for truth \parencite{abeler_preferences_2019}.

%|%\textcite{shaw2011designing} test layperson ability to report content quality correctly, given several incentive schemes. 
\textcite{shaw_designing_2011} test layperson ability to report content quality correctly, given several incentive schemes. 
The researchers find that the intimidation method led to a higher reporting accuracy than proper scoring. The effect that they find is highly significant. 

\textcite{baillon_incentives_2022} study survey responses: The authors employ a randomized trial, where the one group is paid a flat fee (introspection) and the other group is treated with the intimidation method and paid according to the BTS. They find that for questions regarding subjective own health, well-being and language knowledge, there is no difference in reports between the two groups. 
Additionally, the authors conduct the same study with a default-option to induce bias. That is, one of the survey answers is pre-selected. Empirical research shows that default options are more likely to be reported \parencite{johnson_defaults_2003}. In this setting, default answers were also reported significantly more often than in the standard setting. However, the intimidated group showed a slightly lower default bias, i.e. the reports are closer to the unbiased survey setting. This may indicate that respondents exercised more effort and reported more truthfully.

\textcite{lee_testing_2018} test the intimidation method by forecasting the results of NFL games. When filtering reports by respondents who self-identify as \say{extremely knowledgeable}, the predictions perform slightly worse than media experts, which is already not an impressive benchmark.\footnote{The result, as it appears in the original paper, is that the intimidated group performs slightly better than media experts. However, the authors corrected their results in a corrigendum.} The study was replicated and similar results were found in the replicated study \parencite{rutchick_does_2020}.

%|% Weaver and Prelec (2013) experiments 
\textcite{weaver_creating_2013} ask respondents to report truthfully whether they recognized a name. Names included those of historic personas, authors and characters but also technical jargon.
The set of terms shown to respondents also included foil, i.e. made-up names that cannot be recognized. 
The authors run five different randomized trials in which respondents are either in a condition where they are just rewarded with a flat fee (introspection) or rewarded according to their BTS score.
%|% The authors largely do not explain how the BTS works but use the intimidation method 
The authors do not explain how the BTS works but use the intimidation method, except for experiment 3 where no explanation occurred (see next section). 
%|% The general result is that participants who are subject to the intimidation method claim significantly fewer items as recognized and are more accurate in the detection of imaginary items + even when there is a overclaiming bias - The BTS largely cancels this
The result is that respondents who are subject to the intimidation method report significantly fewer items as recognized and are more accurate in the detection of made-up names. 
In certain experiments, bias is introduced by providing respondents with a small monetary reward for reporting recognition of an item, regardless of the accuracy of their report. Thus, respondents who are not intimidated report recognizing significantly more made-up items.
Surprisingly, when combining the incentive to report items as recognized with the intimidation method, the induced bias vanishes almost entirely. The intimidation method is effective in inducing more truthful answers in the presence of a strong incentive to be untruthful.
%|% Experiment 5: subjects state their willingness-to-donate -> Hypothetical (76%) and real donations (44%) are observed (avg.). 
In experiment 5, respondents are asked to report their willingness-to-donate. In one group respondents actually donate their money that they would otherwise receive. In this group, the respondents chose to donate 44\% of their earnings. In contrast, another group whose willingness-to-donate reports were not realized reported that they would donate 77\% of their earnings. This is a clear example of social desirability bias \parencite{krumpal_determinants_2013}.
%|% When using the intimidation method, the subjects tend to report willingness-to-donate closer to the real value (47% and 50% - in two different groups with different BTS experience)
Another group that is subject to the intimidation method reported a willingness-to-donate of 47\%, much closer to the actual donation rate of the first group.  
%|% Again, the intimidation method largely cancels a pre-existing bias
Again, the intimidation method largely mitigates the bias.

%|% The study " A penny for your thoughts" provides largely different results 
\textcite{barrage_penny_2010} ask respondents to collectively vote on whether to donate funds for a public good. If the majority decides in favor of the donation, the donation is realized. 
%|% The researchers set up a randomized trial with students. The study features a "hypothetical" group that was simply asked to state how they would vote, and three additional treatment groups with treatments to induce honesty, among them the intimidation method (with the BTS as the mechanisms). 
The researchers set up a randomized trial with students, which features a \say{hypothetical} group that was told to report how they would vote if the event were real, and three additional treatment groups with treatments to induce honesty, among them the intimidation method (with the BTS as the mechanism). 
%|% The study finds that 40/% of subjects vote to donate and 81\% of subjects state that they would vote to donate if they had to, showing a clear bias.
The study finds that 40\% of respondents vote to donate if donations are realized, whereas 81\% of the respondents in the hypothetical group state that they would vote to donate, showing a clear social desirability bias.\footnote{The study features two votes regarding two different public goods. The answers are averaged here for brevity.} 
%|%The intimidation method does not fully mitigate this bias
The intimidation method does not fully mitigate this bias; 66\% of respondents vote to donate. The intimidation method is as effective in this study as telling respondents about the social desirability bias beforehand.
%|% The study also finds that the intimidation method induces honesty in women and inexperienced subjects only. This is very interesting insofar as it suggests that it the intimidation method may not work well for panels featuring (experienced) males.
The study also finds that the intimidation method induces honesty in women and inexperienced respondents only. This is very interesting insofar as it suggests that the intimidation method may not work well for panels composed of experienced male respondents.\footnote{Furthermore, women display more honesty in behavioral experiments \parencite{abeler_preferences_2019}. The role of gender in honesty is not yet fully understood. This is an avenue for future research.}

\subsubsection{Empirical evidence on the BTS}

%|% On top, the mechanism is not explained (?) to participants in experiment 3 - they learn from their scores only. This is the only actual test of the BTS.
Experiment 3 of \textcite{weaver_creating_2013} poses evidence regarding the effectiveness of the BTS itself because the respondents were \textit{not} informed that the they will be rewarded for honest answers; there is no intimidation statement.
%|% Claim: 
The experiment suggests that the BTS, as conveyed through score feedback, is effective.\footnote{The respondents answered a series of questions in sequence and observed their BTS scores dynamically update after each response.}
%%%% Reason + Evidence:
As in the other experiments in \textcite{weaver_creating_2013}, respondents reported recognizing fewer made-up items as the survey progressed, i.e. reports became more accurate.
Similarly, in Experiment 2 of \textcite{weaver_creating_2013}, respondents—who also saw their BTS scores after each question—appeared to adjust their reporting behavior in response to the feedback, as the accuracy of reports increased as the survey progressed.

%---------------------------------------2.3 SUBSEQUENT RESEARCH ---------------------------------------------------------------%

\subsubsection{Variations of Peer-Prediction and the BTS}\label{sec: Variations of BTS + PP}

%|% BTS and Peer-Prediction are nice, but many mechanism designers have tried to fix some flaws and limitations that were addressed
The papers by \textcite{miller_eliciting_2005} and \textcite{prelec_bayesian_2004} sparked the development of a significant amount of research in the field of mechanism design. Mechanism designers attempted to alleviate some of the obvious practical and theoretical limitations of the two mechanisms.
%%% Claim: +
However, improvements usually come at the cost of increased complexity. This section provides a brief overview. For more technical detail on some mechanisms, see \textcite{faltings_game-theoretic_2023}.

%|% Papers focusing on the large necessary crowd 

\textbf{Crowd size:}
While the original BTS is only applicable for large crowds, \textcite{witkowski_robust_2012} have discovered a version of the BTS that achieves incentive compatibility with only 3 or more respondents. However, this mechanism requires binary reports. \textcite{radanovic_robust_2013} build upon this to achieve a mechanism that works for 3 or more respondents but can handle categorical reports. 

%|% Papers focusing on common priors and posteriors

\textbf{Common priors and posteriors:}
\textcite{witkowski_learning_2013} show that the common prior in Peer-Prediction need not be known if the information of interest is binary. If respondents only make binary reports, the prior can theoretically be estimated from the reports themselves, given a sufficient sample size of reports. The authors also develop a \say{divergence-based-BTS}, which penalizes inconsistency between predictions of others' reports and the own reported belief \parencite{radanovic_incentives_2014, radanovic_incentives_2015}. This mechanism does not require binary reports, large crowds, or common priors.
\textcite{radanovic_incentives_2016} extend the Peer-Prediction mechanism by relaxing the common prior and common posterior assumption. However, this requires multiple questions \parencite{radanovic_incentives_2015}. They call this mechanism the Peer Truth Serum (PTS).
The PTS has also been subjected to preliminary empirical trials \parencite{radanovic_incentives_2016}. \textcite{radanovic_incentives_2016} use a modified version of the mechanism to implement a peer-grading scheme in a real-world class. Similarly, \textcite{timko_incentive_2023} collect self-report data from clickworkers. Both studies use a treatment group in which the \say{basic features} of the mechanism are conveyed. Although the authors check for comprehension, it is not clear whether participants understood the mechanism itself. Both studies find that the mechanism slightly but significantly improves data quality.
\textcite{goel_personalized_2020} show that a version of this mechanism can exploit covariance across multiple questions, which makes it theoretically applicable to purely idiosyncratic estimations, such as incentivizing the honest statement of personal height and gender.

%|% Papers focusing on undesirable equilibria

\textbf{Undesirable Nash equilibria:}
\textcite{dasgupta_crowdsourced_2013} show that the truth-telling equilibrium in the Peer-Prediction mechanism can become \say{focal} in the sense that it becomes the highest-grossing equilibrium for all respondents \parencite{kong_equilibrium_2018}. However, this requires common priors, binary reports, and multiple questions per respondent. These questions also need to be the same across respondents because the mechanism requires panel data. Other works build upon this by relaxing additional assumptions necessary to arrive at this feat \parencite{radanovic_incentives_2016, agarwal_peer_2017}. 
For example, \textcite{shnayder_informed_2016} propose an altered mechanism which they call \textit{Correlated Agreement}. This mechanism still has the benefit of truth-telling being the highest-paying equilibrium, without requiring binary reports as in \textcite{dasgupta_crowdsourced_2013}. The mechanism still requires that each respondent answers multiple questions.

\textcite{kong_information_2019} describe a family of mechanisms with the interesting property that truth-telling is the highest-grossing equilibrium \parencite{kong_dominantly_2024, kong_more_2022}, and in which respondents who are in complete agreement receive no reward. This feature is called \textit{dominant truthfulness}, which is not to be confused with dominant strategies.\footnote{It is impossible to elicit information without verification in dominant strategy as the reward always depends on other respondents \parencite{gao_incentivizing_2020}.}  These mechanisms are very complicated. The proposed mechanism requires that the number of questions is greater than twice the number of possible reports per question. With 10 potential reports per question, 20 questions per respondent would be required. However, the Square-Root-Agreement Rule is less complex and also has many of the desirable properties and fewer requirements \parencite{kamble_square_2023}.

%|% Papers focusing on learning and other behavior

\textbf{Learning and adversarial behavior:}
\textcite{feng_peer_2022} discuss the role of learning in sequential Peer-Prediction. Most of the literature focuses one-shot elicitation, so it is not clear what the effect of repeated interaction is. \textcite{feng_peer_2022} find that the Correlated Agreement mechanism still provides a truthful equilibrium if respondents are assumed to exhibit reward-based learning behavior. \textcite{liu_sequential_2017} also propose a mechanism that can benefit from sequential reporting. Their model also incorporates respondents choice to be effortful but is limited to binary reports.

\textcite{schoenebeck_information_2021} draw upon methods from robust learning to identify adversarial reports.
\textcite{liu_surrogate_2022} study a different class of belief elicitation mechanisms where the principal has access to \textit{some} noisy information on truth. The authors build upon algorithms for unsupervised learning that also need to verify data as correct. They couple this with the use of Peer-Prediction mechanisms to elicit probabilistic beliefs.

\subsubsection{Choice-matching}

%Explain the general mechanism, not the specific version where one uses the prediction frequency

\begin{figure}
    \centering
    \includegraphics[width=.7\textwidth]{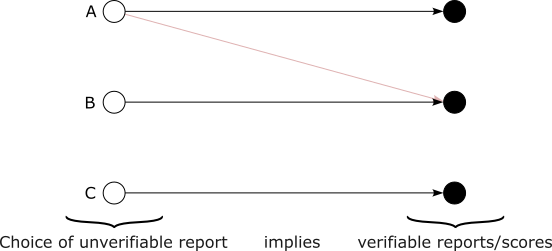}
    \caption{Choice-matching relies on strict correlation between beliefs}
    \label{fig:Choice-matching}
\end{figure}

%|% Claim: 
As most mechanisms covered thus far are highly complex, researchers have sought to provide simpler alternatives. Choice-Matching can be considered a variation of the BTS that is structurally no less complex, but designed to be more intuitive and easier to convey \parencite{cvitanic_honesty_2019}. As in the BTS, the respondents are asked to make two reports: one unverifiable multiple-choice report and a verifiable \textit{auxiliary} report. Respondents are rewarded for accuracy on the verifiable auxiliary report. The answers to the non-verifiable first report sort respondents into groups. Respondents are also rewarded based on the accuracy of auxiliary reports submitted by other members of their group. Specifically, each respondent’s reward is calculated as a weighted sum of their own score and the average group score. For the verifiable auxiliary report truthful reporting is incentivized through a proper scoring rule, making honesty the dominant strategy. For the non-verifiable report truthfulness is a BNE if respondents believe that those who share their own belief about the non-verifiable report also make the most accurate verifiable reports, and if everyone reports truthfully.
Essentially, choice-matching substitutes the scoring of an unverifiable report (see Figure \ref{fig:Choice-matching}) by inferring an associated verifiable report.
 Consider the purely illustrative question: \say{What percentage of global energy will come from fusion in the year 2100?} The auxiliary question is: \say{Will a company have reached a near-term fusion energy milestone in 2030?} A fusion-skeptic should report a low probability of the fusion-milestone being reached. If the fusion-skeptic reports a low percentage of fusion-energy by 2100 this effectively assigns her a reward on the auxiliary (fusion-milestone) question from other respondents who also report a low fusion-energy percentage. This report is likely to better match her own belief than the reports from respondents who report a higher percentage of fusion by 2100. Therefore, she is better off by being truthful.

The key assumption is that respondents assume a continuous correlation between the reported beliefs on both questions. If this assumption is not met, incentives can be dramatically different.  For example, if belief A in Figure \ref{fig:Choice-matching} is believed to correspond to the same implied verifiable report as B (red arrow), there is no difference between reporting A or B for the respondents. 

The study by \textcite{zawojska_incentivizing_2022} shows that the real-world implementation of choice-matching is not straightforward. \textcite{zawojska_incentivizing_2022} use choice-matching to elicit the willingness-to-pay (WTP) of the inhabitants of Warsaw regarding new solar panels. These solar panels are a hypothetical intervention of the city of Warsaw. The second auxiliary question is: \say{Given that you have 1000 zloty, how much of this budget are you going to spend on reforestation?} The key assumption here is that respondents' preferences regarding reforestation and WTP for solar panels are strictly correlated. Respondents were informed that the true donation towards reforestation will be determined by their own reported WTP for solar panels and the reports of others who stated a similar WTP. Assuming that the mechanism works, stating a high WTP for solar panels would imply a high donation to reforestation and vice versa.
The study showcases the following issues: 
\begin{enumerate}
    \item The choice of the auxiliary question is not straightforward. A strict correlation of beliefs is required, which is not satisfied in the study by \textcite{zawojska_incentivizing_2022}. They run a survey to find out that only 54\% of participants believed in a positive correlation and as much as 25\% of respondents believed WTP for solar panels and reforestation to be negatively correlated.
    \item This study features an additional complication involving donations, as respondents retain any funds they choose not to donate. This creates an opportunity for strategic manipulation: a respondent interested in maximizing overall donations could report a low WTP for solar panels, thereby being grouped with others expected to make smaller donations to reforestation. However, this respondent could donate a lot to reforestation (auxiliary report). This would artificially increase the average donation in the low-WTP/low-donation group. The respondent could then donate all funds that she received to reforestation, effectively maximizing charitable donations. Had the respondent been honest, her entire share of funds would have gone to reforestation, but the donations in the low-WTP group would be lower.
    \item Choice-matching requires categorical reports in order to \say{group} respondents. This potentially sacrifices the accuracy of reports. There were only seven choices in the study by \textcite{zawojska_incentivizing_2022}.\footnote{In this study, seven groups were constructed \textit{post-hoc} based on the elicitation of WTP with numerical values. This way, full accuracy is maintained. However, this post-hoc grouping (through the experimenter) determines the outcomes of all respondents and may thus be seen as an unfair and unwanted feature.} 
\end{enumerate}

When faced with choice-matching, respondents reported a significantly higher average WTP for solar energy when compared with a control group that is not incentivized. Another interesting finding from the study is that the introduction of choice-matching impedes self-reported understanding. Since the study cannot compare the outcomes with observable measures of WTP for solar energy, it does not provide evidence for or against choice-matching.

\subsubsection{Square Root Agreement Rule}

The Square Root Agreement Rule attempts to improve upon Output Agreement by making truth-telling a BNE which is the highest-grossing equilibrium. The difference to Output Agreement is that respondents are not simply given a fixed reward if their answers match. Rather, a fixed reward is divided by the square root of a popularity index that indicates how many others have made the same report. In other words, the Square Root Agreement Rule incentivizes matching reports that are rare in the general population. It requires common priors, multiple questions, and a sufficient sample size.
To find out which report is highest-grossing, the respondent should first estimate what other responses are likely to look like. The highest-grossing report coincides with the respondent's true belief regarding the question, given a set of assumptions. An obvious downside is that the probability of matching a random peer is small with many similar but distinct reports. For example, consider a probability elicitation: \say{What is the probability of X happening?}. If two matched peers report 9\% and 8\% respectively, they get no reward although they are almost agreeing. \textcite{kong_dominantly_2024} criticizes the Square Root Agreement Rule for having a comparably \say{weak truthfulness property}, as the truthful equilibrium is only guaranteed to be higher-grossing if all respondents are truthful across all questions.

\subsubsection{Source differential peer prediction}

\begin{figure}
    \centering
    \includegraphics[width=0.7\textwidth]{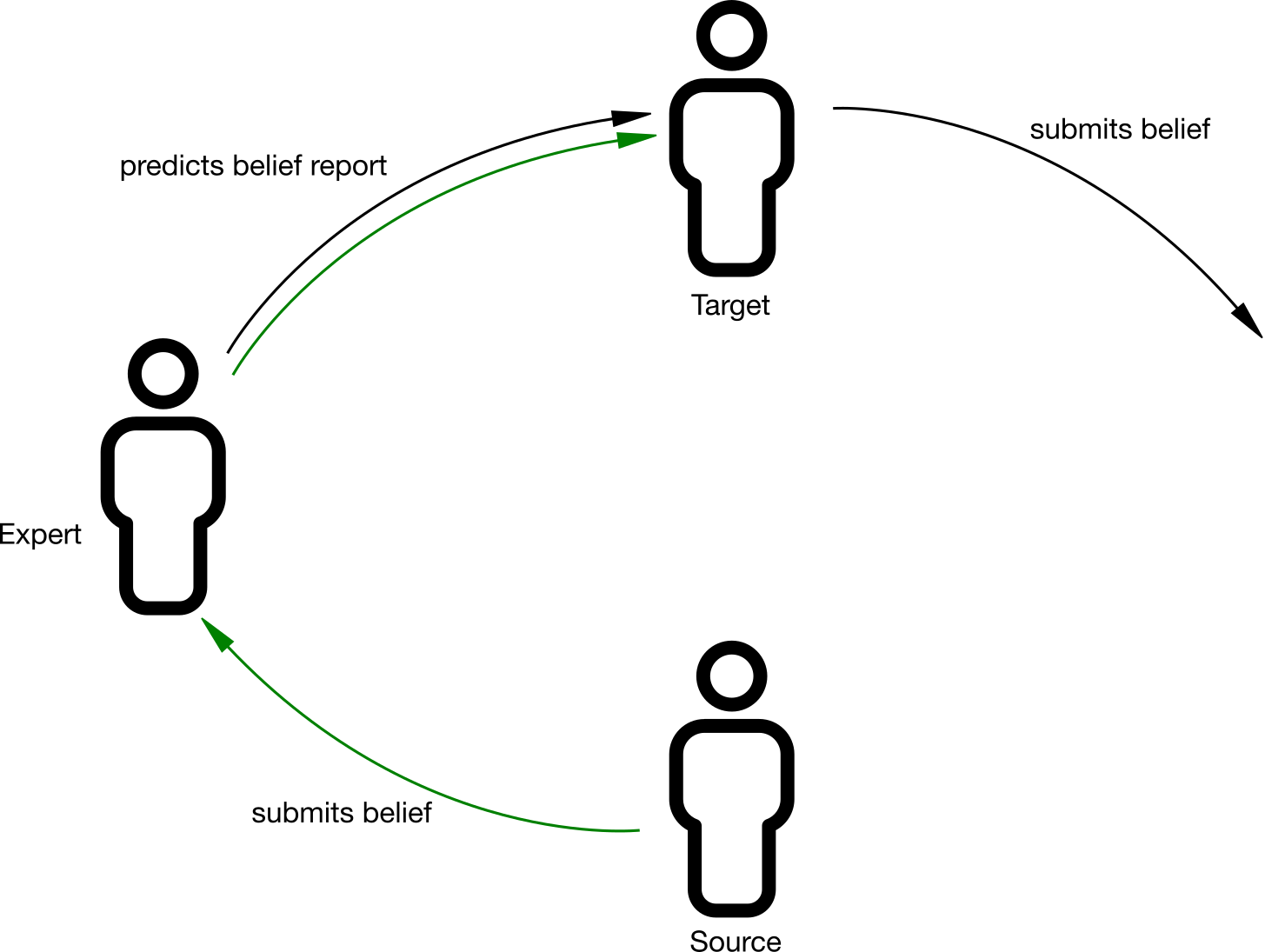}
    \caption{In source differential peer prediction the expert makes two predictions of the targets prediction, one before and one after the sources prediction has been unveiled to the expert.}
    \label{fig:Source_differential_peer_prediction}
\end{figure}

%%% Claim: 
Source differential peer prediction is an intuitive mechanism that assigns respondents different roles \parencite{schoenebeck_two_2020}.\footnote{Actually, the authors propose two mechanisms. The other mechanism is called \textit{target differential peer prediction}. The two are not described separately here because they are two versions of the same mechanism.} The roles are expert, source and target (see Figure \ref{fig:Source_differential_peer_prediction}).\footnote{Actually, the expert does know that she is the expert, and which role the others have. However, leaving the expert uncertain regarding her role is possible. \textcite{srinivasan_auctions_2021} shows this in a hypothetical application. This is an avenue for future research.} 
All respondents report their true beliefs simultaneously.
The expert predicts the target's report and is rewarded based on the accuracy of her prediction. The target is not rewarded for accuracy or honesty. The source's report is handed to the expert \textit{after} the expert made her prediction. The expert then \textit{revises} her prediction and predicts the targets report a second time. Both the expert and the source are rewarded based on the accuracy of this revised report. What are the incentives set up via this game? The expert is best off accurately predicting the targets signal. However, no one knows who is source or target. Thus, the expert is incentivized to predict the average report. Since source and target do not know which role they have, they have the same incentives. If they are assigned the role of the source, they will maximize their rewards if the expert makes the best possible revised prediction. Therefore, the source/target is incentivized to supply the expert with the best possible information. This theoretically results in both target and source reporting their true beliefs. Truth-telling is the highest-grossing BNE in this game.

%----------------------- Market-based mechanisms ---------------------%

\subsection{Market-based mechanisms}\label{sec: market based mechanisms}

%|% Some researchers have taken a different route and have employed market-based mechanisms for IEWV. Although similar, these offer some advantages over mechanisms that are based on reports.

%Bayesian Market

\subsubsection{Bayesian Market}

On a \textit{Bayesian Market}, respondents purchase assets and consequently reveal their true beliefs \parencite{baillon_bayesian_2017}. 
 %|% Claim: 
The Bayesian Market reduces complexity for respondents relative to most other mechanisms because 
%%%% Reason:
respondents have a simple decision to make: buy or sell.
The asset's fundamental value on a Bayesian Market is the relative frequency of 'Yes' or 'Buy' positions. That is, if 60\% of the respondents choose to purchase the asset, the value of the asset is \$0.60. The respondents can either take a 'Yes' or 'No' position, i.e. buy or sell (short) the asset. All trading occurs via the market maker and not directly between respondents. The Bayesian \say{Market} does not resemble a real market because the market maker fully controls the price. The Bayesian market is just a simple bet. A hypothetical example best illustrates the Bayesian Market. A hypothetical panel of respondents is asked: \say{Will the company fail?} Let the market maker set a price of \$0.35.\footnote{The price of the asset could also be determined randomly. However, the principal might already know which price is sensible, if he can guess the common prior.} If a respondent believes that more than 35\% of respondents will purchase, then it is profitable to purchase oneself. If less than 35\% of respondents take a 'Yes' position, the asset is not worth its price. It is profitable to take a 'No' position. The market maker facilitates the transactions at a strictly positive cost to himself, and in turn receives a frequency of 'Yes' positions. From these, the respondents' true beliefs are deductible if one assumes the CPSS case. In the example, the common prior could be that 30\% of the respondents believe that the company is going to fail. Respondents who received information that the company is going to fail update their beliefs and assume that more than 35\% of others also believe the company to fail. They will take 'Yes' positions. Respondents who received contrary information will take a 'No' position. Truth-telling is a BNE.

%%%% Claim: 
The Bayesian Market seems like a very promising candidate for actual application, but it can only elicit \textit{binary} reports. The issue can be resolved by running multiple Bayesian Markets. Every probability distribution can be approximated by discrete versions, which could be elicited with binary questions. This would result in questions like: \say{Do you think that the probability with which humanity is going to face a major crisis because of AI is at least x\% this century (Yes/No)?} Having 100 questions of this kind theoretically allows one to elicit probabilistic beliefs more accurately. However, this would be extremely cumbersome, impractical, and tiring for respondents.\footnote{\textcite{atanasov_improving_2025} find that eliciting beliefs using what they call \say{menus}—lists of binary choices—requires respondents to spend twice as much time as compared to other interfaces, such as text boxes. Additionally, they observe that respondent retention over the survey period is significantly lower with menus than with other elicitation methods.}

\subsubsection{Peer betting and variations of the Bayesian Market}

%%%% Claim: Peer betting is an extension of the Bayesian Market, quite similar and pretty cool
Peer betting is very similar to the Bayesian Market and can be seen as an extension. \textcite{baillon_peer_2025} make important advances in the modeling of belief elicitation
%%%% Reason: Because the authors discuss a more realistic setting 
because the authors incorporate effort, adversarial incentives, and preferences for truth into the model \parencite{baillon_follow_2025}. Each of these is incorporated as a cost to be paid by the respondent. 
%%% Claim: 
This leads to notable observations of its own.
\textcite{baillon_peer_2025} find that the Peer betting mechanism can incentivize truthfulness and effort in the face of adversarial incentives, and that increasing the rewards to respondents strictly increases their truthfulness. They also find that adversarial incentives lead to reduced effort; that is, respondents are less likely to acquire informative signals.
%%%
\textcite{baillon_peer_2025} also relax the common prior assumption, but implement a barely weaker common prior expectations assumption.

% More simple bets to elicit private beliefs

\textcite{baillon_simple_2021} further develop the Bayesian Market by showing that simple bets can reveal private signals without the need for common priors, or for agents to agree when receiving the same information. However, the mechanisms outlined in \textcite{baillon_simple_2021} are more complicated and still only applicable to the elicitation of binary reports.
%%% Claim: 
\textcite{baillon_follow_2025} amend the Bayesian Market by individualizing the prices/bets offered to respondents. This does not affect the incentives that each individual respondent faces, but \textcite{baillon_follow_2025} claim that it has more desirable properties with respect to aggregation.

\subsubsection{Empirical evidence on Peer betting and Bayesian Markets}

%%%% Claim: It works, but one study does not find an effect
The peer betting mechanism is effective in preliminary empirical trials, increasing effort and truth-telling \parencite{baillon_peer_2025}.
Unlike other empirical trials, the analysis by \textcite{baillon_peer_2025} actually tests the mechanism itself.
%%% Reason: Participants understand incentives 
Most notably, the rewards are completely explained to respondents. The respondents do not even get the hint that truth-telling is best or an equilibrium strategy.
%%% Evidence: 
The authors conduct a randomized trial in which respondents need to pick the \say{correct} box out of two. There are three groups: Introspection (control), Peer betting, and actual bets (verified outcome). The \say{true} box is predetermined and known to the experimenter; the outcome is verifiable. The boxes each have marbles of blue and yellow color in them. The respondents know the sum of blue and the sum of yellow marbles across the two boxes as well as a the minimum amount of either yellow or blue marbles in each box.
This gives the respondents minimal but important information on the potential distributions of marbles across the two boxes. The respondents are then faced with a choice. At any point, respondents can select a box, reporting their belief that it is the true one. Alternatively, they can choose to draw a marble from the true box.\footnote{The marble is placed back into the box. Drawing the marbles does not affect the contents of the boxes.} To draw the marble and make this valuable observation, respondents have to take an additional task that is not rewarded, engaging in effort to update their own beliefs.  The main result of the study is that the three treatments induce significantly different levels of effort. Additional effort is most often observable in the group which makes real bets, where respondents are directly rewarded if they guess the box correctly. The Peer betting group engaged in fewer additional tasks, showing less effort. However, this group did engage in additional tasks significantly more often than the introspection group. Prediction accuracy was positively affected by engaging in effort. The study's results support the plausible causal chain: 

\begin{center}
\begin{math}
    \text{Incentives} \rightarrow \text{Effort} \rightarrow \text{Additional information} \rightarrow \text{Prediction Accuracy}
\end{math}
\end{center}

Thus, this study provides an extremely valuable insight: respondents performed prediction tasks in absence of verification with greater accuracy when incentivized, because incentives motivated them to gather additional information.

Furthermore, in a second study by \textcite{baillon_peer_2025}, Peer betting leads to a 10\% increase in self-reported violations of safety guidance related to infectious disease. Given that violation of safety guidance may be perceived as antisocial, this result suggests that the Peer betting incentives (partially) overcome a social desirability bias.

%%% Acknowledgement and Response: 
However, Peer betting also reduced self-reported understanding in the second study of \textcite{baillon_peer_2025}. Furthermore, a Bayesian market treatment did not affect truthful reports relative to introspection in a separate study \parencite{baillon_follow_2025}. Given that the Bayesian market treatment is so similar to Peer betting, additional research is necessary to help us understand these results.

\subsubsection{Self-resolving prediction markets}

% Prediction Market for unresolvable outcomes

On \textit{self-resolving prediction markets}, respondents buy or sell an asset \textit{sequentially}, as on an actual prediction market \parencite{srinivasan_self-resolving_2023}. All trading occurs via a market maker, which also makes self-resolving prediction markets a misnomer, as it is more of a sequential Peer-Prediction mechanism than a traditional market. The value of the asset is equal to the closing price. The market terminates at any trade with a pre-specified probability. Essentially, respondents try to predict the closing value of the asset. \textcite{srinivasan_self-resolving_2023} prove that, if everyone reports truthfully, truth-telling is a BNE in approximation.

%Self-resolving Information Markets (Theory - Empirical analysis should come later)

This raises the question: Why not use a conventional prediction market? Belief elicitation and aggregation on markets and prediction markets is well studied \parencite{koessler_information_2012, arrow_promise_2008}.\footnote{For example, bond prices predict economic growth and recessions \parencite{estrella_how_2003}. Betting odds on sports games accurately predict game outcomes \parencite{spann_sports_2009}.} 
This idea has been put forward \parencite{ahlstrom-vij_self-resolving_2020, slamka_secondgeneration_2012}. \textcite{ahlstrom-vij_self-resolving_2020} proposes to use a market that terminates at a random point in time and where the fundamental value of an asset is the closing price.\footnote{Regular prediction markets can also be self-resolved in case evaluation becomes difficult. The Many Labs 2 study, a large-scale replication project for psychology research, was taken as an opportunity to test how well psychologists can predict replication outcomes. \textcite{forsell_predicting_2019} set up a prediction market that ended up correctly predicting 75\% of replication outcomes. Because the replications took far longer than anticipated, the researchers decided to resolve the prediction markets based on the final prices. No participating psychologist objected to this, which suggests that the participants seemed to find the self-resolution fair. This provides no evidence in favor of self-resolving markets, as participants believed the market to be a regular prediction market when they made predictions.} \textcite{ahlstrom-vij_self-resolving_2020} argues that in the absence of any clear equilibrium strategy, reporting truthfully may be a focal strategy, because humans prefer to be truthful \parencite{abeler_preferences_2019}. However, as such a market constitutes a bubble without a fundamental value, there is no other reason to believe that self-resolving information markets will be effective.

\subsubsection{Empirical evidence on self-resolving prediction markets}

\textcite{ahlstrom-vij_self-resolving_2020} conducts a randomized trial. One group is assigned to play on self-resolving prediction markets, whereas the control group plays on a regular prediction market. The respondents acted with play-money only. The most successful trader got a small bonus payment on top of the reward for participating. The predictive accuracy of both markets was compared to determine whether self-resolving prediction markets are significantly worse. Respondents were tasked with predicting the share of black balls in an urn. The respondents individually observed independent draws from the urn over time and the evolving price on the market. All respondents observed the same market price, thus receiving indirect information about other respondents' observations. Surprisingly, self-resolving prediction markets were as accurate as regular prediction markets in predicting the share of black balls. This finding is also highly statistically significant. However, the result must be interpreted with respect to the study's setting. Respondents were recruited via Prolific and had a shockingly low rate of passing the comprehension check. Perhaps they did not understand how the self-resolving prediction market differs from a regular prediction market. This would explain why there is no difference between the two markets. More generally, it is uncertain how this 10-minute Prolific experience generalizes to settings we care about.

\textcite{slamka_secondgeneration_2012} also conducted an experiment to investigate prediction markets that do not resolve based on outcomes. \textcite{slamka_secondgeneration_2012} compare three different proposed designs: A market that self-resolves randomly in time, a market that self-resolves at a fixed date, and a market whose asset's value is the volume-weighted average market price across time. In the case of the first two markets, the fundamental asset value is the closing price. The authors run an experiment with an additional control group that trades on a regular prediction market. The study design is similar to the one by \textcite{ahlstrom-vij_self-resolving_2020}, but with two additional treatments. However, the total sample size is much smaller (N=78), leading to small treatment groups. The result is that all treatments, i.e. self-resolving markets, performed slightly worse than the regular market. Since prediction markets are widely recognized to be one of the most effective methods of eliciting accurate beliefs from a crowd \parencite{arrow_promise_2008}, both the study by \textcite{ahlstrom-vij_self-resolving_2020} and \textcite{slamka_secondgeneration_2012} set a high bar for their self-resolving prediction markets (see also section \ref{sec: Benchmarking mechanisms}). Worrisome for this study is not the difference in predictive accuracy between treatment and control, but the absolute error displayed by all groups. All three treatment groups performed no better than chance on binary questions.\footnote{The reported t-value implies a 36\% chance of achieving a more accurate forecast than the \textit{regular} prediction market provided by adopting random guessing as a strategy.} It is highly questionable whether respondents actually had insight at all into the outcomes of binary questions. This greatly dilutes the result that otherwise would have been largely favorable regarding the application of self-resolving prediction markets.

%--------------------------------------- 3 APPLICATIONS ---------------------------------------------------------------%

\section{Applied research} \label{sec: Applied research}

IEWV mechanisms have been used to elicit truthful reports regarding questionable research practices, criminal conduct, long-term energy price forecasting, experimental philosophy, and other topics.
%|%  Claim: Applications of IEWV mechanisms do not provide any evidence that they work
These studies do not provide any evidence of the effectiveness of certain mechanisms 
%|% Reason: Because we cannot compare answers to objective probabilities or beliefs elicited in some other way. 
because we cannot compare answers to objective truth.

%%%% Claim: 
However, we observe that the application of IEWV mechanisms proves challenging, 
%%%% Reason: 
as many researchers struggle to implement mechanisms coherently.
%%% Evidence: 
\textcite{john_measuring_2012} try to estimate how many psychologists engage in questionable research practices and outright fraud. The authors conduct a randomized trial, in which one group is subject to the intimidation method, i.e. the authors link to the paper by \textcite{prelec_bayesian_2004} and assure participants that the BTS \say{rewards truthful answers}. The control group is assigned to introspection. Since respondents were anonymized, payments were instead made to charities on behalf of the respondents. The respondents were asked the following questions for ten different questionable research practices: 

\begin{enumerate}
    \item Did you cheat? (private prediction/self-admission)
    \item What percentage of your peers cheated? (Peer-Prediction/prevalence)
    \item How many of your cheating peers will admit to cheating?
\end{enumerate}

The first report is idiosyncratic, i.e. refers to the \textit{own} engagement in questionable research practices. The BTS still incentivizes truthfulness on such an idiosyncratic report if one understands the own engagement in questionable research practices as the private signal that updates the prevalence (Peer-Prediction) of questionable research practices from a common prior. This implies that the expected prevalence of questionable research practices is assumed to be \textit{exclusively} informed by the own behavior.
%|% To illustrate: How about if someone wanted to learn about the height of a certain population without measuring a subsample: This person asks the entirety (or a representative subsample of the population): What is your height? What is the distribution of height in the population? The BTS is used to score individuals. This would work out in theory, if everyone had the same perception of height in the population that is additionally only informed by the own height. However, people tend to perceive others heights too. In the CPSS, signals are private. Height is not private. Thus, beliefs might be formed based on the heights of people that are around (signals of others).
This is a bold assumption. Furthermore, the authors ask for a third report that seeks to estimate how many psychologists will report untruthfully and not admit to questionable research practices in the survey. The BTS only incentivizes truth-telling if everyone else is truth-telling. Given that the authors assumed this would be violated, the BTS cannot have been incentivizing truthful reporting. Yet, the authors did not inform respondents about this.

The data show that telling psychologists that they are rewarded with the BTS increases self-admission rates of questionable research practices slightly. Even psychology researchers seem to fall for the intimidation method. The study reports shockingly high numbers of questionable research practices.
\Textcite{van_de_schoot_use_2021} replicate the study by \textcite{john_measuring_2012}, finding similar results.

In a similar study, \textcite{loughran_incentivizing_2014} investigate the effectiveness of the intimidation method for self-reporting criminal conducts and misdemeanours. The respondents were asked questions such as:

\begin{enumerate}
    \item Did you engage in drunk driving? 
    \item What percentage of your peers engage in drunk driving?
\end{enumerate}

As in the study by \textcite{john_measuring_2012}, \textcite{loughran_incentivizing_2014} compute BTS scores from these two questions, treating the report to the first question as the respondent-specific signal and the second as the Peer-Prediction. Although this is an interesting application, it is unlikely that key assumptions for the BTS to work properly are met because the experimenters assume that the own behavior exclusively informs prevalence estimates and that everyone shares a common prior, which is questionable. The authors deployed the intimidation method and did not explain the BTS. The result of the study is that respondents assigned to the intimidation method reported a higher willingness to offend.

% + The forecasts are terribly imprecisely stated, allowing no useful judgment regarding their accuracy or the forecasters skill 
\textcite{zhou_long-term_2019} study the effectiveness of the BTS for long-term energy price forecasts. However, the paper has many methodological flaws. The authors elicit binned forecasts regarding future energy price changes from experts. That is, the experts report in which range their expectation lies. This makes sense as they are using the BTS to reward truthful reports, and the number of potential reports needs to be limited. However, the number of choices could arguably have been larger. The authors only provide 7 possible choices, which relate to price changes in percent relative to 2015 prices.\footnote{The intervals are: $(-\infty;-16],[-15;-8],[-7;-3],[-2;2][3;7][8;15][16;\infty)$} The ranges are very large, have varying width, and are open to negative and positive infinity, making interpretation of the forecasts difficult \parencite{kruger_quantifying_2024}. The authors do not use proper scores or any other error measure for the observed outcomes and simply state the short-term forecasts (for which outcomes are available) \say{accurately predicted} outcomes. Given the absence of error measures or forecasting benchmarks, this statement lacks any meaning. Overall, the study completely ignored state-of-the-art forecasting techniques. The study does provide an interesting result in its own way, which the authors fail to discuss: The scores for different reports, as calculated by the BTS, are non-steadily decreasing across possible reports. An expert that predicts oil prices to increase by 51\% or more  in Australia would have received a high positive score. An expert who predicted an oil price increase of 11\% to 25\% would have too received a high positive score. However, an expert who predicted an oil price increase of 26\% to 50\% would have gotten a negative score. This simply makes no sense.\footnote{A forecast of 26-50\% cannot be worse than both the predictions of 11-25\% and 51\%. This also implies that assumptions regarding the BTS are violated because the BTS revives the logarithmic scoring rule, which is proper and single-peaked, when assumptions are met.} Furthermore, the authors show the experts three videos outlining different future scenarios prior to eliciting beliefs. These videos are clearly a form of conceptual cues and may have affected reports \parencite{weingarten_primed_2016}. The authors carried out another study, using \textit{the same} scenarios to \say{stretch the thinking of survey participants and to help focus their minds [...]} to better forecast car-sharing market penetration \parencite{zhou_projected_2017}, but do not discuss or study the effects of these cues. 

%\textcite{goel_infochain_2020} propose to use IEWV mechanisms for providing information to blockchain-based applications. Whilst blockchain technology essentially tries to circumvent the need for trusted third parties, whenever a real-world variable is needed to resolve smart contracts, which are simply digital contracts on the blockchain, such as whether shipment of an order occurred, access to truth is necessary. \textcite{goel_infochain_2020} suggest to use IEWV mechanisms to incentivize truth-telling in a crowd of non-partisan individuals. Thus, the crowd becomes the trusted third party that judges over the smart contract. 
 
\textcite{schoenegger_experimental_2023} proposes to use incentives in experimental philosophy. Incentivization in experimental philosophy is not straightforward as ground truth is exactly the subject to be debated. However, most IEWV mechanisms are built around the assumption that respondents perceive a noisy signal from a \textit{shared} observable truth. Whether this is the case in areas of philosophy is not clear. The author conducted a randomized controlled trial where one group of respondents received a fixed reward for participation and one group was rewarded with the BTS. The study too deployed the intimidation method. Respondents were sourced from Prolific. The study finds that telling respondents that they will be scored better, if they report truthfully, changes responses on four out of seven philosophical questions significantly. However, a follow-up investigation by \textcite{schoenegger_taking_2022} produced a null result. The second study employed different questions, albeit not completely different topics. Therefore, it is hard to attribute causes for the different results. Overall, the studies' results still support the notion that changing incentives and using the intimidation method affects responses, but the results warrant closer examination and more future research is needed to disentangle the effects of incentives on reports in experimental philosophy.

%\textcite{srinivasan_auctions_2021} propose a mechanism to improve academic peer review. Firstly, all papers submitted to an outlet shall accompany a bet which signals the willingness-to-pay for a review. The editors decision to desk-reject is replaced by a Vickrey-Clarke-Groves auction (\cite{groves_incentives_1973}). That is, there are a limited number of papers that can be handed in for peer review, and the authors that make the highest bets win those slots. The authors pay the price that the highest non-reviewed (auction-losing) paper specified. In the review stage, each paper is reviewed by three reviewers who each submit a report. The reviewers are paid according to a mechanism similar to the source-differential Peer-Prediction mechanism (\textcite{schoenebeck_two_2020}). Thus, the funds raised via the first auction stage are used to pay reviewers based on how well their review predicts other reviews \textit{and} improves predictions of other reviews, thereby incentivizing honest and effortful reviews. 

Output Agreement mechanisms have been used in experimental research to elicit unverifiable opinions from respondents. A brief discussion of these studies can be found in Section 2.2.4 of \textcite{charness_experimental_2021}.

%---------------------

%What are some applications you can think of that would fit well for IEWV methods? 

%----- Forecast combination ------%

\section{Forecast combination}\label{sec: forecast_combination}

%--------------------------------------------------
%Nomenclature: 

% Aggregation/Weighting/combination = forecast combination

% Forecasts/reports/Submitted beliefs = forecasts

%------------------------------------------------

%|% IEWV mechanisms are being repurposed to work for forecast aggregation. Although that is no longer IEWV, it is still covered here, because the literature is so adjacent.

%|% COntext:
Some of the mechanisms for IEWV have been repurposed for combining or aggregating forecasts. Forecast combination refers to distilling a single estimate from multiple estimates. Simple combination schemes, such as taking the mean, have been found to robustly improve upon individual forecasts \parencite{clemen_combining_1989}. It is important to remark that forecast combination is a post-hoc analysis and is unrelated to the \textit{elicitation} of beliefs.
%|% Claim: 
IEWV mechanisms seem to be a promising way to combine forecasts
%%% Reason: 
because they can help emphasize shared information or discover skilled forecasters in the crowd.
%%%Acknowledgement & Response
However, some combination methods do not outperform simpler alternatives in empirical tests.

\textcite{prelec_solution_2017} propose the \say{Suprisingly popular} (SP) algorithm, which shares its Bayesian framework with the BTS. This algorithm employs the Peer-Prediction as a second question. The  relative performance of forecasters is assessed using this second question, assuming that accurate forecasters also make accurate Peer-Predictions. This information is then used to combine forecasts. \textcite{prelec_solution_2017} conduct empirical studies across various domains and respondent groups, using the SP algorithm to combine forecasts. These forecasts are then compared to majority opinion and confidence-weighted combination schemes. The result is that the SP combination scheme provides the correct answer more often than other combination schemes. However, the chosen benchmarks are not state-of-the-art combination schemes.\footnote{None of the benchmark combination schemes is reviewed in the extensive reviews by \textcite{clemen_combining_1989} or \textcite{wang_forecast_2023}. Whilst \textcite{prelec_solution_2017} aim to fix majority voting, majority voting is not usually considered as a way of combining forecasts.} Moreover, even the mean predictions are not much better than chance on all datasets and individual predictions are not much better than chance \parencite{liu_surrogate_2022}.\footnote{The Brier scores of means are between 0.333 and 0.480 \parencite{wang_forecast_2021}. The mean predictions are almost certainly better than individual predictions. The high Brier scores come from the fact that forecasters made Yes/No statements.}
Furthermore, the SP does not improve significantly upon majority voting or confidence-weighted combinations in the study by \textcite{wilkening_hidden_2022} and did not perform much better than simpler alternatives in three other studies \parencite{rutchick_does_2020}.\footnote{\textcite{wilkening_hidden_2022} also replicate the study by \textcite{prelec_solution_2017} and find the results to be robust. Respondents are barely better than a random guesser in the studies by \textcite{rutchick_does_2020} so that there is little to wisdom to combine.}
\textcite{radas_whose_2019} find in successive randomized trials that combining predictions via the SP reduces error in reported willingness to pay for new gadgets.

A problem is that the study by \textcite{prelec_solution_2017} only considers categorical reports, i.e. no probabilistic predictions are being made. It would be standard forecasting practice to make probabilistic predictions, thus expressing uncertainty \parencite{gneiting_probabilistic_2014}.
%%%% Claim: Probability forecasts can be used with other mechanisms 
Other variations of IEWV mechanisms have been developed into combination methods that can handle probabilistic information 
%|% Evidence: 
\parencite{martinie_using_2020, peker_extracting_2023, rilling_neutral_2024, palley_extracting_2019}.
%%% Claim: Combination Algorithms can discover forecasters in the crowd 
IEWV mechanisms can discover skilled forecasters in the crowd 
%%% Reason: 
because the scores should be closely related to proper scores \parencite{prelec_bayesian_2004}. In other words, we can use the scores from IEWV mechanisms as a proxy for proper scores as long as the latter are not yet available because the event has not yet occurred.

\begin{figure}
    \centering
    \includegraphics{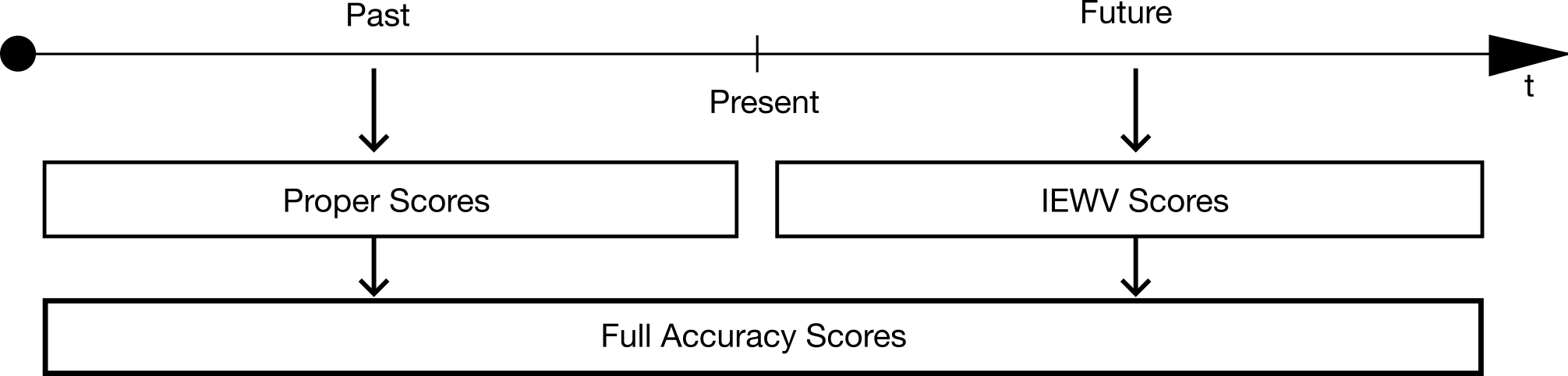}
    \caption{Full Accuracy Scoring leverages information from predictions about future and past events}
    \label{fig:FAS}
\end{figure}

\textcite{atanasov_full_2023} propose \textit{Full Accuracy Scoring}, in which yet-to-be-verified forecasts are scored using IEWV methods, while resolved forecasts are evaluated with proper scoring rules.
Thus, individual forecasters (or models) are evaluated in terms of their \say{full accuracy} as opposed to their \say{past accuracy} (see Figure \ref{fig:FAS}). 

\textcite{atanasov_full_2023} largely focus on discovering skilled forecasters in a forecasting tournament.
The authors find that forecasting skill, as measured by proper scores after all questions have resolved, is indeed better predicted by Full Accuracy Scoring than past track-records at any point in time, when computing Full Accuracy Scores on two past forecasting tournaments.\footnote{\textcite{atanasov_full_2023} report results for two different forecasting datasets that they employ in their empirical analysis. Proper Proxy Scores are barely predicting proper scores in study 1. They predict proper scores in study 2 extremely well, outperforming the past track record for much of the time. Why is that? The paper lacks an explanation for this observation and other key differences between the two datasets.} The authors chose Proper Proxy Scoring Rules as the IEWV method for scoring yet-to-be-evaluated forecasts.  However, the advantage of Full Accuracy Scoring diminishes as the share of events that have happened in the past increases. This means that Full Accuracy Scoring is most useful when the track record of forecasters is small.
Full Accuracy Scoring seems to be particularly interesting for improving the accuracy of long-term-forecasts.

\textcite{wang_forecast_2021} also find that Peer-Prediction mechanisms may be discovering skilled forecasters in a crowd. An important difference between Peer-Prediction mechanisms and the SP is that Peer-Prediction mechanisms usually do not require that forecasters report the Peer-Prediction. Therefore, Peer-Prediction mechanisms can be used to combine single forecasts when a common prior is assumed. 
This has the additional benefit that the methods can be tested on historical forecasting data. The authors use Surrogate Scoring Rules, the Peer Truth Serum, Proxy Scoring Rules, Determinant based Mutual Information, and Correlated Agreement to rank forecasters in historical forecasting datasets. The bottom 90\% of forecasters, as measured by the Peer-Prediction scores, are eliminated and only the remaining forecasts are used for further forecast combination. Surprisingly, forecast combination with all five Peer-Prediction algorithms slightly improve total accuracy in many datasets. Particularly, accuracy in the GoodJudgment Forecasting Project is improved, even compared to a very tough benchmark \parencite{satopaa_combining_2014}.  All five Peer-Prediction mechanisms achieve very similar performance in combining the forecasts.

\textcite{martinie_using_2020} propose a combination scheme to identify confident forecasters. As with the SP method, respondents report the own belief and the Peer-Prediction. The forecasts are then weighted by the difference between the own prediction and the prediction of others prediction. Intuitively, this mechanism should identify forecasters who possess a lot of private information. \textcite{wilkening_hidden_2022} find that a variation of this combination scheme, called \say{surprisingly confident}, is significantly better than the SP in combining predictions.

%%%% Claim: IEWV mechanisms can solve the shared information problem 

\textcite{palley_extracting_2019} motivate their forecast combination scheme, called \say{pivoting}, by explaining how it solves the \say{shared-information-problem}. 
%%% Reason:
Generally, forecast combination seems to improve upon individual forecasts because it aggregates diverse information that different forecasters possess \parencite{clemen_combining_1989}. However, if multiple forecasters have shared information, this information is over-emphasized when taking the mean of forecasts. That is, forecasters that are similar to each other do not provide additional value but bias the combined forecast. By posing the additional question: \say{What will your peers forecast?}, combination methods can \textit{in theory} detect forecasters with shared information. This is of course only the case if the assumptions regarding the beliefs of forecasters are correct. The authors propose multiple weighting schemes based on different scenarios regarding the theoretical information structure. 
%%%% Evidence:
\textcite{palley_extracting_2019} find in four empirical studies that their combination method shows potential benefits. However, the studies also show that the superiority of the combination method fades as the prediction task becomes less lab-controlled and closer to real-world forecasting. In studies 1 and 2, the authors set up a very abstract scenario that matches the theoretical assumptions well. Respondents recruited via MTurk are asked to predict the bias in a biased coin based on a limited number of private and public observations. In these studies the combination methods improve upon the mean estimate slightly but significantly. In study 3, students are asked to predict the price of groceries. In this setting, all methods again improve upon simple averaging, but the improvement is much smaller than in the studies 1 and 2. In study 4, respondents recruited via MTurk forecast NCAA basketball game outcomes, where the forecast combination methods reduce error only by approximately one percent compared to the mean. Furthermore, there are two factors that may have impacted the results of study 3 and 4. First, students in study 3 did not actually purchase groceries and were not otherwise scored for accuracy. Respondents in study 4 had poor knowledge of basketball game outcomes, barely beating chance.\footnote{The average Brier score was $0.232$. The average Brier score of a random guesser would have been $0.25$.} 

%%% Claim: 

Recent studies have achieved further mild improvements in combining predictions using shared information. 
%%%%Evidence
The combination methods by \textcite{peker_extracting_2023} , \textcite{palley_boosting_2023} and \textcite{martinie_using_2020} improve Brier scores in a set of general knowledge questions.\footnote{See \textcite{peker_extracting_2023} section 5.}
\textcite{rilling_neutral_2024} develops another variation that outperforms the methods by \textcite{palley_extracting_2019}, \textcite{peker_extracting_2023} and \textcite{palley_boosting_2023} in 8 out of 12 datasets.

%%%% Claim: Although several combination schemes improve upon alternatives, benefits fade as the prediction task becomes less lab-controlled and closer to real-world forecasting.
Although these combination methods improve upon alternatives, benefits seem to heavily depend on the topic. For example, none of the combination schemes significantly improves upon the mean when forecasting NCAA Basketball games \parencite{rilling_neutral_2024}, and there is substantial variance in error reduction across datasets.

Although there is both rationale and evidence to suggest that the combination methods discussed in this section are valuable, further benchmarking against other state-of-the-art combination methods is needed.

%---------------------------------------5 CRITICISM and FUTURE RESEARCH---------------------------------------------------------------%
\section{Discussion and future research directions}\label{sec: Discussion}

\subsection{From theory to practice: The need for empirical research}

%%%% Claim + Evidence: Choosing mechanisms is a fundamentally empirical matter. 
%|% to predict real play, we need empirical research.
\textbf{Since we are interested in \say{what works}, comparing mechanisms is a fundamentally empirical matter \parencite{charness_experimental_2021}}. 
%%%% Reason: There are limits to what theory can tell us.
Whilst theory is useful for finding IEWV mechanisms, there are limits to its ability to discern their effectiveness because there is no theoretical property that guarantees strong incentives to be truthful, and there is no fully collusion-resistant mechanism.  Respondents can always collude to maximize their rewards.\footnote{Even mechanisms that are entirely built around the idea of avoiding this, such as in \textcite{kong_dominantly_2024}, cannot distinguish between true reports and collective permutations of it (such as everyone reporting the opposite).}
%%%% Claim: Furthermore, we have reason to doubt that theoretical assumptions are not violated in reality
Furthermore, theoretical assumptions are likely to be violated to varying degrees in reality
%%%% Reason: Because they are usually quite strict.
because they are usually quite strict and specific.
Almost all mechanisms assume the CPSS, a Bayesian update from a common prior based on a single signal, or a similar version of this belief formation process. However, this assumption is only valid in situations where participants simply relay a signal, such as in data-labeling or self-report surveys; it does not hold in more complex settings.

%%%% Claim: However, we observe a research focus on theory over empirical research. 
\textbf{Despite the importance of empirical research,} we observe a focus on theory over empirics.
%%%% Evidence: As evidenced via this review. 
This review article documents that empirical research has been gravely neglected. There are only a few mechanisms for which some empirical evidence has been collected.
%%%% Reason: A reason for this is that empirical research is more difficult to do (see next section)
We conjecture that this is because theoretical research has driven mechanism complexity, which impedes empirical tests of mechanisms.
%|% Background: Why mechanisms are so complex
 %Claim: Mechanism Designers have neglected the comprehension.
\textbf{Mechanism designers have neglected complexity in pursuit of more strongly truthful mechanisms.} %Evidence:
Mechanisms that were published later are often more complicated, but tend to have more \say{desirable} theoretical properties. Mechanism designers are keenly aware of the challenge that elicitation may involve large numbers of (potentially unmotivated) respondents with limited attention. Unfortunately, designing mechanisms to be 'detail-free' and require minimal input is misguided, as it has usually resulted in mechanisms that are actually more complex and harder to understand.

%Claim: Many mechanisms are very complex, and this impedes research and application.
\textbf{The complexity of mechanisms impedes research and application.}
%Warrant: Why empirical research requires that subjects understand mechanisms
It is important that respondents understand how mechanisms work.
If there is no way to convey the mechanism to respondents, it cannot have any effect on their behavior.  
%|% Reason : Complex mechanisms cannot be tested. If they cannot be tested, they cannot be assumed to work well in practice. 
Consequently, if a mechanism is too complicated to be understood by respondents, it cannot be tested.
%%%% Evidence: 
We not only see evidence for this problem in the lack of empirical research, but also in the existing empirical research.
%|% Background: The state of empirical research 
Most empirical studies have resorted to measures that avoid explanation of the actual mechanism. These studies consequently fail to provide convincing evidence that mechanisms incentivize truthfulness. 

%|% The use of forecasting data for "verification" of mechanisms is a mistake. 
Furthermore, many authors try to verify their mechanism empirically \say{with real human forecasters}, using synthetic forecasting data to compare proper scores with scores computed with IEWV mechanisms, claiming success if the two match. This is a fruitless exercise and not informative because the whole point of IEWV is to change incentives and behavior in the first place. If the forecasters had not been incentivized with proper scoring, forecasts might have been different. We would not need any mechanism if we knew that reports are already truthful.

%|% Solution: Make mechanisms easier to perceive and understand
\textbf{The solution is clear: Make mechanisms easier to understand.} Interestingly, this is also the bottom line of \textcite{charness_experimental_2021}, who review elicitation methods generally and thus reach the same conclusion using different evidence. More specifically, they conclude that there is a trade-off between complexity and theoretical appeal of the elicitation method and \say{that the tendency of researchers
studying belief elicitation to design ever-more theoretically-robust methods with little consideration of complexity has not led to systematic improvements in empirical belief measurements}. This holds true for mechanisms studied in this review.
%|% Theoretical research has discovered intuitive mechanisms, which suggests that it may be possible to have mechanism that can be understood intuitively. 
Recent research suggests that the search for easy-to-comprehend and truthful mechanisms is far from over, and that significant improvements can be made. Choice-matching is much more intuitive than the BTS, and the Bayesian Market is easier to explain than Peer-Prediction, all whilst mostly retaining desired theoretical properties.

\subsection{Selection and incentivization of respondents in empirical research}

%|% another reason why studies failed to provide. Convincing evidence of the effectiveness of mechanisms are the subject themselves.
%|% Claim: Crowdscourcing platforms are regularly used for sourcing participants, but participants show poor knowledge regarding outcomes
\textbf{Empirical studies often fail to convincingly demonstrate the effectiveness of mechanisms, because respondents frequently display little foresight—a limitation that impedes empirical research.}
%|% Warrant:
If respondents have no foresight, then there is nothing to study, as treatment will not measurably affect the accuracy of reports.
%|% Evidence - and reason:
This review remarks no fewer than six times that study participants perform barely better than a random guesser.
%|% There is good reason to believe that crowdworkers may be poor forecasters. This requires additional research. 
There is good reason to believe that crowdworkers, who constitute most research respondents, may be poor forecasters and estimators. For example, some studies ask respondents to make probabilistic predictions, although we know that untrained respondents tend to make poor probabilistic statements due to miscalibration \parencite{jungermann_calibration_1977, keren_facing_1987}.
Expertise in the subject matter and the ability to make accurate forecasts typically varies considerably across respondents \parencite{tetlock_superforecasting_2015}.
%Claim: 
%|% This requires additional research. Take experts, or at least calibrated forecasters, or at least students.
Thus, additional research that investigates the aptitude of crowdsourcing platforms for this kind of research would be valuable.
Sourcing respondents from online forecasting tournaments or universities may be better and more representative regarding applications involving expert elicitation \parencite{tetlock_forecasting_2014}. 

%|%Paying above thresholds is not incentive compatible %

\textbf{Many of the reviewed studies do not stick to the mechanisms that they test, but add lotteries or bonus payouts, which sabotages the incentive structure.}
Most empirical studies pay only top respondents a fixed bonus, e.g. the top $1/3$ of respondents, as determined by IEWV scores. This massively distorts the original incentives of the mechanism, undermining any considerations regarding truthful play along the way \parencite{witkowski_incentive-compatible_2023, ottaviani_strategy_2006}. This is a mostly theoretical remark since the mechanisms themselves are rarely understood by participants. However, researchers should pay attention to this in future empirical work.

\subsection{Benchmarking mechanisms with introspection}\label{sec: Benchmarking mechanisms}

%|% In practice introspection is King! And this is the benchmark to beat, not proper scoring! 
\textbf{Introspection is the benchmark to beat, not proper scoring.} A lot of the focus on truthful equilibria stems from the misguided sense of having to recreate the incentives for truth-telling as provided by proper scoring. \textcite{prelec_bayesian_2004} motivates the BTS by showing that it recreates the incentives of the logarithmic scoring rule. This would be great, but it is not necessary. When ground truth is not accessible, proper scoring is not available. There should be no doubt that introspection is currently the method that is used to elicit beliefs when ground truth is inaccessible, and is the standard in research practice outside of economics \parencite{schoenegger_experimental_2023, charness_experimental_2021}. 
%|% Claim:
Furthermore, introspection mostly does just fine \parencite{charness_experimental_2021, baillon_incentives_2022, dana_are_2019}. %Reason + Evidence
%|% IEWV mechanisms mustnt make humans more truthful, but more helpful! Any improvement over introspection is welcome
 IEWV mechanisms are \textit{helpful} if they lead to more accurate reports than introspection. Output Agreement, as studied by \textcite{von_ahn_labeling_2004}, shows that even the simplest mechanism can potentially improve upon introspection. 

\textbf{Mechanisms can also be harmful by distracting and incentivizing collusion.} The study by \textcite{gao_trick_2014} shows that this may be a very real problem. \textcite{charness_experimental_2021} argue that the complexity of mechanisms reduces the accuracy of reports because it confuses respondents. Furthermore, the use of mechanisms may discourage effort by interfering with the experts intrinsic motivation to answer accurately \parencite{frey_motivation_2001}.
%|% we should be careful with implementing mechanisms in reality - much fieldwork needs to be done
Mechanisms need to be applied with great care and only after additional fieldwork is done to make sure that mechanisms do not pose adversarial incentives.

\subsection{Future research directions}\label{sec: most important open research questions}

%|% We need simple mechanisms with intuitive incentives for providing accurate estimates. 

%|% The gold standard for answering the question of whether IEWV can be employed as a direct decision support would be to ran a randomized trial with skilled forecasters, some subject to a IEWV scoring mechanism. Their task would be to predict the results (causal effects) of not-yet-published but pre-registered studies. This would provide additional evidence of counterfactual judgment over simple prediction. It may not be prudent to run this study without any prior knowledge of which mechanism would be well suited. Thus, some empirical groundwork should be done first. 

\subsubsection{Open theoretical research questions}

\begin{itemize}
    \item Are there simpler mechanisms in which truthfulness constitutes a BNE?

    How can such mechanisms be made more robust to collusion, or modified so that the incentives are more intuitive? 

    \item Assumptions regarding the respondents' belief formation process need to be made. Are they met? How can we know?

    Assumptions regarding the respondents' belief formation process make testable predictions about the elicited data. For example, a common prior and impersonal updating imply that two respondents with the same Peer-Prediction must have the same private prediction and vice versa. This can be tested post-hoc. Since verification is possible in studies (to see whether mechanisms have an effect on accuracy), the scores of a hypothetical additional respondent can be analyzed. They can be compared to the proper scores to see whether mechanisms do revive them, as theoretically promised. If the incentives are not proper, assumptions are likely to be violated.\footnote{See e.g. the discussion of the study by \textcite{zhou_long-term_2019} in section \ref{sec: Applied research}.}

    \item What is the role of risk preferences?

    Most articles implicitly assume that respondents are risk-neutral utility maximizers, but this assumption is rarely examined. If respondents instead exhibit different risk preferences, it raises the important question of whether the mechanisms remain incentive-compatible under these conditions.

\end{itemize}

\subsubsection{Open empirical and experimental research questions}

%|% Claim: 
Convincing evidence on the effectiveness of mechanisms can only come from randomized trials because they allow to compare the mechanisms with introspection.
% Given the thin layer of empirical evidence, more basic empirical analysis is necessary first. Smaller trials of promising mechanism inform which mechanisms should be tested on a larger scale. 
Given the thin layer of empirical evidence, more basic empirical analysis is necessary first. Smaller trials of promising mechanisms can inform which mechanisms should be tested on a larger scale. There are three key considerations that apply to most empirical work: 

\begin{enumerate}
    \item \textit{Perception} of incentives affects behavior, not the actual mechanism. Since the mechanisms need to be explained to the participating respondents, the actual treatment is the explanation and how respondents perceive the mechanism based on that explanation. Perception can be tested by creating two equally correct explanations of a mechanism and test whether they actually yield the same behavior. 

    \item Perception may depend very much on the respondent. A mechanism that helps to improve accuracy in inexpert self-reports does not necessarily improve the accuracy of expert judgment. Different mechanisms may be needed for different applications.

    \item A second path to affect perception of the incentives is \textit{repeated interaction}. Respondents can learn about the mechanisms through past rewards related to their responses, and the study by \textcite{weaver_creating_2013} suggests that this can be a powerful path to affect perception.
\end{enumerate}

Open research questions include:

\begin{itemize}

    \item Are respondents effortful? 

    Experiments can test whether the incentives induce the respondents to engage in more effort. A good example of this is the study by \textcite{baillon_peer_2025}, where respondents can engage in additional tasks to obtain more signals.

    \item How well do self-resolving information markets work in practice?

    Self-resolving information markets look great in existing empirical studies. Empirical studies that are closer to real-world application are needed. A simple first step could be to test self-resolving information markets on existing play-money markets. Self-resolving markets could be created there, and the accuracy compared to chance. This would help determine whether more experienced traders in these markets are able to exploit or manipulate self-resolving information markets.

    %/% What is the actual question here? More research necessary!
    \item What is the role of the false-consensus effect?

    \textcite{carvalho_inducing_2017} argues that a false-consensus-effect, (falsely) believing that the own opinion is the majority opinion, explains why Output Agreement is effective. Is the false-consensus-effect really the cause? Is the false-consensus-effect rational when agents update based on their information, as is exploited in Peer-Prediction and Truth Serums? 

    \item Why do mechanisms (not) work? 

    Experiments should distinguish between different ways in which the instructions might affect truth-telling behavior of respondents. Respondents may be unconditionally honest \parencite{abeler_preferences_2019}, actually react to incentives posed, or blindly trust a claim that truth-telling is in their own best interest. Furthermore, it could be that the pure act of being held accountable for reports influences behavior, or that this exacerbates experimenter demand effects \parencite{lerner_accounting_1999, zizzo_experimenter_2010}.

    \item How collusion-resistant are different mechanisms?

    Since respondents rewards only depend on their reports, respondents can always collude to report whatever maximizes their rewards, regardless of truth. There is no mechanism that can completely avoid this. Thus, a key property of mechanisms is to make collusion as difficult as possible. The collusion-resistance of mechanisms can be experimentally verified by giving groups a question or task, controlled access to mechanisms and the opportunity to collude on answers. If collusion is straightforward, the group will collude to get a higher reward. If collusion is more difficult, the group might find that answering the question honestly is the easier way to receive a high reward.

    \item Are there ways to improve upon Proper Proxy Scoring? Which proxies would provide the strongest incentives to be accurate?

    Should we take the simple mean? Or should we resort to extremizing \parencite{baron_two_2014} or other complicated proxies that work well in forecast combination?

    \item Who is truthful? Within-subject design vs. treatment groups.

      After respondents answer a question which is incentivized with an IEWV mechanism, respondents could be offered a fair bet against the true outcome. If the betting decision is not in line with the first answer, this can be interpreted as evidence of being non-truthful in the first stage.

    \item What is the effect of the intimidation method across studies? Should we employ the intimidation method when it is credible? 

    The treatment effect of the intimidation method could be assessed by gathering data from multiple studies. The more important question is: when and whether to use the intimidation method? It could do great harm to science and the credibility of research to make false claims, even if they are well-intended. Ethical considerations are necessary.

    \item Are crowdsourcing platforms a good source of respondents for studies on forecasting and IEWV? 

    By gathering the data from the studies that are reviewed here, and studies on forecasting that employed crowdworkers, the accuracy of crowdworkers on these judgment tasks could be assessed. A randomized trial involving respondents from multiple sources could assess differences in performance.

    \item  Does Full Accuracy Scoring meaningfully change existing long-term forecasts?

    Since full accuracy scoring appears to improve forecasts, it would be valuable to combine long-term forecasts—whether made by models or individuals—using information from short-term forecasts made by the same sources. 

    \item Do respondents believe key assumptions to be met? 
    
    We can simply ask respondents whether they believe the key assumptions are met, as is done in some of the reviewed studies when assessing self-reported understanding. All mechanisms require that all respondents are truthful and expect others to be truthful too. Eliciting the expected truthfulness of other respondents is thus very interesting, see e.g. \textcite{weaver_creating_2013}.

    \item What is the role of gender in belief elicitation?

    \textcite{barrage_penny_2010} find that the intimidation method induces honesty in women to a much greater extent than in others. Similarly, \textcite{abeler_preferences_2019} finds that women resist incentives to misreport stronger, i.e. have stronger preferences for honesty. Overall, these findings suggest that there are gender differences in responsiveness to incentives. However, these differences are not yet well understood.

\end{itemize}

%--------------------------------------- 6 CONCLUSION---------------------------------------------------------------%
\section{Conclusion}\label{sec: Conclusion}

%|% The main point: This review article summarized research related to mechanisms which incentivize truth-telling in absence of verification.
This review article provides an overview of mechanisms which incentivize truth-telling in absence of verification.
%|% Significance: The significance of these methods has not diminished. We are reliant on accurate and truthful reports for sound decision-making in many areas, such as e.g. estimating existential risk.
Accurate and unverifiable self-reports are crucial in areas such as long-term forecasting, estimating risk, and data labeling.

%|% The look back: Despite the fact that much theoretical research has been done and many mechanisms have been proposed, we lack empirical evidence
Although there has been extensive theoretical research into mechanisms that provide incentives for truth-telling in the absence of verification, this article shows that empirical evidence is scarce.
%\% Why we lack empirical evidence 
It is difficult to gather empirical evidence on the effectiveness of mechanisms because their complexity makes it hard to convey them to respondents. 
%\% Therefore, simpler mechanisms could be more valuable.
Therefore, devising and testing simpler mechanisms is an important avenue for future research.

%|% Open research questions: 
Furthermore, this review suggests that many research questions remain to be solved. We do not yet understand when or if to use mechanisms, whether assumptions are violated, what the role of cognitive biases is, where to source research participants, and how these mechanisms could be implemented at scale. 
%|% I am looking forward to future randomized trials with calibrated subjects that provide convincing evidence of the mechanisms potential to elicit beliefs even when ground truth is inaccessible.
Developing tools to more accurately elicit beliefs when the truth cannot be verified remains an important goal for future research.

%\nolinenumbers

\printbibliography

@article{baillon_peer_2025,
	title = {Peer betting to elicit unveriﬁable information},
	url = {https://aurelienbaillon.com/research/papers/pdf/peerbetting.pdf#page=1.71},
	language = {en},
	author = {Baillon, Aurelien and Peker, Cem and van der Zee, Sophie},
	month = feb,
	year = {2025},
}

@incollection{jungermann_calibration_1977,
	address = {Dordrecht},
	title = {Calibration of {Probabilities}: {The} {State} of the {Art}},
	isbn = {978-94-010-1278-2},
	shorttitle = {Calibration of {Probabilities}},
	url = {http://link.springer.com/10.1007/978-94-010-1276-8_19},
	language = {en},
	urldate = {2024-09-05},
	booktitle = {Decision {Making} and {Change} in {Human} {Affairs}},
	publisher = {Springer Netherlands},
	author = {Lichtenstein, Sarah and Fischhoff, Baruch and Phillips, Lawrence D.},
	editor = {Jungermann, Helmut and De Zeeuw, Gerard},
	year = {1977},
	doi = {10.1007/978-94-010-1276-8_19},
	pages = {275--324},
}

@incollection{schoenebeck_two_2020,
	address = {Cham},
	title = {Two {Strongly} {Truthful} {Mechanisms} for {Three} {Heterogeneous} {Agents} {Answering} {One} {Question}},
	volume = {12495},
	isbn = {978-3-030-64945-6},
	url = {http://link.springer.com/10.1007/978-3-030-64946-3_9},
	language = {en},
	urldate = {2024-06-13},
	booktitle = {Web and {Internet} {Economics}},
	publisher = {Springer International Publishing},
	author = {Schoenebeck, Grant and Yu, Fang-Yi},
	year = {2020},
	doi = {10.1007/978-3-030-64946-3_9},
	note = {Series Title: Lecture Notes in Computer Science},
	pages = {119--132},
}

@book{law_human_2011,
	address = {Cham},
	series = {Synthesis {Lectures} on {Artificial} {Intelligence} and {Machine} {Learning}},
	title = {Human {Computation}},
	copyright = {https://www.springer.com/tdm},
	isbn = {978-3-031-00427-8},
	url = {https://link.springer.com/10.1007/978-3-031-01555-7},
	language = {en},
	urldate = {2024-09-05},
	publisher = {Springer International Publishing},
	author = {Law, Edith and Von Ahn, Luis},
	year = {2011},
	doi = {10.1007/978-3-031-01555-7},
}

@book{faltings_game_2017,
	address = {Cham},
	series = {Synthesis {Lectures} on {Artificial} {Intelligence} and {Machine} {Learning}},
	title = {Game {Theory} for {Data} {Science}: {Eliciting} {Truthful} {Information}},
	copyright = {https://www.springer.com/tdm},
	isbn = {978-3-031-00449-0},
	shorttitle = {Game {Theory} for {Data} {Science}},
	url = {https://link.springer.com/10.1007/978-3-031-01577-9},
	language = {en},
	urldate = {2024-06-13},
	publisher = {Springer International Publishing},
	author = {Faltings, Boi and Radanovic, Goran},
	year = {2017},
	doi = {10.1007/978-3-031-01577-9},
}

@book{parmigiani_decision_2009,
	address = {Chichester, West Sussex},
	title = {Decision theory: principles and approaches},
	isbn = {978-0-471-49657-1},
	shorttitle = {Decision theory},
	language = {eng},
	publisher = {John Wiley \& Sons},
	author = {Parmigiani, G.},
	collaborator = {Inoue, Lurdes Y. T. and Lopez, Hedibert Freitas},
	year = {2009},
}

@article{ottaviani_strategy_2006,
	title = {The strategy of professional forecasting},
	volume = {81},
	copyright = {https://www.elsevier.com/tdm/userlicense/1.0/},
	issn = {0304405X},
	url = {https://linkinghub.elsevier.com/retrieve/pii/S0304405X06000286},
	doi = {10.1016/j.jfineco.2005.08.002},
	language = {en},
	number = {2},
	urldate = {2025-04-25},
	journal = {Journal of Financial Economics},
	author = {Ottaviani, M and Sorensen, P},
	month = aug,
	year = {2006},
	pages = {441--466},
}

@article{lerner_accounting_1999,
	title = {Accounting for the effects of accountability},
	volume = {125},
	number = {2},
	journal = {Psychological bulletin},
	author = {Lerner, Jennifer S and Tetlock, Philip E},
	year = {1999},
	note = {Publisher: American Psychological Association},
	pages = {255},
}

@article{arrow_promise_2008,
	title = {The {Promise} of {Prediction} {Markets}},
	doi = {10.1126/science.1157679},
	journal = {Science},
	author = {Arrow, Kenneth J. and Forsythe, Robert and Gorham, Michael and Hahn, Robert W. and Hanson, Robin and Ledyard, John O. and Levmore, Saul and Litan, Robert E. and Milgrom, Paul and Nelson, Forrest D. and Neumann, George R. and Ottaviani, Marco and Schelling, Thomas C. and Shiller, Robert J. and Smith, Vernon L. and Snowberg, Erik and Sunstein, Cass R. and Tetlock, Paul C. and Tetlock, Philip E. and Varian, Hal R. and Wolfers, Justin and Zitzewitz, Eric},
	year = {2008},
}

@article{rilling_neutral_2024,
	title = {Neutral {Pivoting}: {Strong} {Bias} {Correction} for {Shared} {Information}},
	issn = {1545-8490, 1545-8504},
	shorttitle = {Neutral {Pivoting}},
	url = {https://pubsonline.informs.org/doi/10.1287/deca.2024.0227},
	doi = {10.1287/deca.2024.0227},
	language = {en},
	urldate = {2025-03-19},
	journal = {Decision Analysis},
	author = {Rilling, Joseph},
	month = dec,
	year = {2024},
	pages = {deca.2024.0227},
}

@article{peker_extracting_2023,
	title = {Extracting the collective wisdom in probabilistic judgments},
	volume = {94},
	issn = {0040-5833, 1573-7187},
	url = {https://link.springer.com/10.1007/s11238-022-09899-4},
	doi = {10.1007/s11238-022-09899-4},
	language = {en},
	number = {3},
	urldate = {2025-03-19},
	journal = {Theory and Decision},
	author = {Peker, Cem},
	month = apr,
	year = {2023},
	pages = {467--501},
}

@misc{baillon_follow_2025,
	title = {Follow the money, not the majority: {A} mechanism predicting unresolvable events},
	url = {https://aurelienbaillon.com/research/papers/pdf/Follow_the_money.pdf},
	language = {en},
	author = {Baillon, Aurelien and Tereick, Benjamin and Wang, Tong V},
	month = jan,
	year = {2025},
}

@misc{cvitanic_honest_2024,
	title = {Honest {Binary} {Choice}: {The} {Two} {Player} {Case}},
	shorttitle = {Honest {Binary} {Choice}},
	url = {https://www.ssrn.com/abstract=5030732},
	doi = {10.2139/ssrn.5030732},
	urldate = {2025-03-17},
	author = {Cvitanic, Jaksa and Prelec, Drazen and Radas, Sonja and Sikic, Hrvoje},
	year = {2024},
}

@misc{atanasov_improving_2025,
	title = {Improving {Low}-{Probability} {Judgments}},
	url = {https://www.ssrn.com/abstract=5025990},
	doi = {10.2139/ssrn.5025990},
	urldate = {2025-03-03},
	author = {Atanasov, Pavel D. and Consigny, Coralie and Karger, Ezra and Schoenegger, Philipp and Budescu, David V. and Tetlock, Philip},
	year = {2025},
}

@article{koessler_information_2012,
	title = {Information aggregation and belief elicitation in experimental parimutuel betting markets},
	volume = {83},
	copyright = {https://www.elsevier.com/tdm/userlicense/1.0/},
	issn = {01672681},
	url = {https://linkinghub.elsevier.com/retrieve/pii/S0167268112000406},
	doi = {10.1016/j.jebo.2012.02.017},
	language = {en},
	number = {2},
	urldate = {2025-02-14},
	journal = {Journal of Economic Behavior \& Organization},
	author = {Koessler, Frédéric and Noussair, Charles and Ziegelmeyer, Anthony},
	month = jul,
	year = {2012},
	pages = {195--208},
}

@article{ahlstrom-vij_self-resolving_2020,
	title = {Self-resolving {Information} {Markets}: {A} {Comparative} {Study}},
	volume = {13},
	issn = {1750-676X},
	shorttitle = {Self-resolving {Information} {Markets}},
	url = {http://www.ubplj.org/index.php/jpm/article/view/1687},
	doi = {10.5750/jpm.v13i1.1687},
	language = {english},
	number = {1},
	urldate = {2024-06-11},
	journal = {The Journal of Prediction Markets},
	author = {Ahlstrom-Vij, Kristoffer},
	year = {2020},
}

@article{offerman_truth_2009,
	title = {A {Truth} {Serum} for {Non}-{Bayesians}: {Correcting} {Proper} {Scoring} {Rules} for {Risk} {Attitudes}},
	volume = {76},
	issn = {00346527, 1467937X},
	shorttitle = {A {Truth} {Serum} for {Non}-{Bayesians}},
	url = {https://academic.oup.com/restud/article-lookup/doi/10.1111/j.1467-937X.2009.00557.x},
	doi = {10.1111/j.1467-937X.2009.00557.x},
	language = {en},
	number = {4},
	urldate = {2024-12-11},
	journal = {Review of Economic Studies},
	author = {Offerman, Theo and Sonnemans, Joep and Van De Kuilen, Gijs and Wakker, Peter P.},
	month = oct,
	year = {2009},
	pages = {1461--1489},
}

@article{hossain_binarized_2013,
	title = {The {Binarized} {Scoring} {Rule}},
	volume = {80},
	issn = {0034-6527, 1467-937X},
	url = {https://academic.oup.com/restud/article-lookup/doi/10.1093/restud/rdt006},
	doi = {10.1093/restud/rdt006},
	language = {en},
	number = {3},
	urldate = {2024-12-05},
	journal = {The Review of Economic Studies},
	author = {Hossain, T. and Okui, R.},
	month = jul,
	year = {2013},
	pages = {984--1001},
}

@misc{greaves_research_2020,
	address = {University of Oxford},
	title = {A research agenda for the {Global} {Priorities} {Institute}},
	url = {https://globalprioritiesinstitute.org/wp-content/uploads/GPI-research-agenda-version-2.1.pdf},
	author = {Greaves, Hilary and MacAskill, William and O’Keeffe-O’Donovan, Rossa and Trammell, Philip and Tereick, Benjamin and Mogensen, Andreas and Tarsney, Christian and Alexandrie, Gustav and Sévricourt, Maxime Cugnon},
	year = {2020},
}

@article{abeler_preferences_2019,
	title = {Preferences for {Truth}‐{Telling}},
	volume = {87},
	issn = {0012-9682},
	url = {https://www.econometricsociety.org/doi/10.3982/ECTA14673},
	doi = {10.3982/ECTA14673},
	language = {en},
	number = {4},
	urldate = {2024-10-28},
	journal = {Econometrica},
	author = {Abeler, Johannes and Nosenzo, Daniele and Raymond, Collin},
	year = {2019},
	pages = {1115--1153},
}

@article{barrage_penny_2010,
	title = {A penny for your thoughts: {Inducing} truth-telling in stated preference elicitation},
	volume = {106},
	copyright = {https://www.elsevier.com/tdm/userlicense/1.0/},
	issn = {01651765},
	shorttitle = {A penny for your thoughts},
	url = {https://linkinghub.elsevier.com/retrieve/pii/S0165176509003681},
	doi = {10.1016/j.econlet.2009.11.006},
	language = {en},
	number = {2},
	urldate = {2024-09-11},
	journal = {Economics Letters},
	author = {Barrage, Lint and Lee, Min Sok},
	month = feb,
	year = {2010},
	pages = {140--142},
}

@article{timko_incentive_2023,
	title = {Incentive {Mechanism} {Design} for {Responsible} {Data} {Governance}: {A} {Large}-scale {Field} {Experiment}},
	volume = {15},
	issn = {1936-1955, 1936-1963},
	shorttitle = {Incentive {Mechanism} {Design} for {Responsible} {Data} {Governance}},
	url = {https://dl.acm.org/doi/10.1145/3592617},
	doi = {10.1145/3592617},
	language = {en},
	number = {2},
	urldate = {2024-09-09},
	journal = {Journal of Data and Information Quality},
	author = {Timko, Christina and Niederstadt, Malte and Goel, Naman and Faltings, Boi},
	month = jun,
	year = {2023},
	pages = {1--18},
}

@inproceedings{faltings_game-theoretic_2023,
	address = {Macau, SAR China},
	title = {Game-theoretic {Mechanisms} for {Eliciting} {Accurate} {Information}},
	isbn = {978-1-956792-03-4},
	url = {https://www.ijcai.org/proceedings/2023/740},
	doi = {10.24963/ijcai.2023/740},
	language = {en},
	urldate = {2024-09-09},
	booktitle = {Proceedings of the {Thirty}-{Second} {International} {Joint} {Conference} on {Artificial} {Intelligence}},
	publisher = {International Joint Conferences on Artificial Intelligence Organization},
	author = {Faltings, Boi},
	month = aug,
	year = {2023},
	pages = {6601--6609},
}

@article{zizzo_experimenter_2010,
	title = {Experimenter demand effects in economic experiments},
	volume = {13},
	copyright = {http://www.springer.com/tdm},
	issn = {1386-4157, 1573-6938},
	url = {http://link.springer.com/10.1007/s10683-009-9230-z},
	doi = {10.1007/s10683-009-9230-z},
	language = {en},
	number = {1},
	urldate = {2024-09-05},
	journal = {Experimental Economics},
	author = {Zizzo, Daniel John},
	month = mar,
	year = {2010},
	pages = {75--98},
}

@article{weingarten_primed_2016,
	title = {From primed concepts to action: {A} meta-analysis of the behavioral effects of incidentally presented words.},
	volume = {142},
	issn = {1939-1455, 0033-2909},
	shorttitle = {From primed concepts to action},
	url = {http://doi.apa.org/getdoi.cfm?doi=10.1037/bul0000030},
	doi = {10.1037/bul0000030},
	language = {en},
	number = {5},
	urldate = {2024-09-05},
	journal = {Psychological Bulletin},
	author = {Weingarten, Evan and Chen, Qijia and McAdams, Maxwell and Yi, Jessica and Hepler, Justin and Albarracín, Dolores},
	month = may,
	year = {2016},
	pages = {472--497},
}

@article{ungar_good_2012,
	title = {The {Good} {Judgment} {Project}: {A} {Large} {Scale} {Test} of {Different} {Methods} of {Combining} {Expert} {Predictions}},
	language = {en},
	journal = {AAAI Technical Report FS-12-06},
	author = {Ungar, Lyle and Mellors, Barb and Satopää, Ville and Baron, Jon and Tetlock, Phil and Ramos, Jaime and Swift, Sam},
	year = {2012},
}

@book{tetlock_superforecasting_2015,
	address = {New York},
	edition = {First edition},
	title = {Superforecasting: the art and science of prediction},
	isbn = {978-0-8041-3669-3},
	shorttitle = {Superforecasting},
	language = {eng},
	publisher = {Crown Publishers},
	author = {Tetlock, Philip E. and Gardner, Dan},
	year = {2015},
}

@article{tetlock_forecasting_2014,
	title = {Forecasting {Tournaments}: {Tools} for {Increasing} {Transparency} and {Improving} the {Quality} of {Debate}},
	volume = {23},
	issn = {0963-7214, 1467-8721},
	shorttitle = {Forecasting {Tournaments}},
	url = {https://journals.sagepub.com/doi/10.1177/0963721414534257},
	doi = {10.1177/0963721414534257},
	language = {en},
	number = {4},
	urldate = {2024-09-05},
	journal = {Current Directions in Psychological Science},
	author = {Tetlock, Philip E. and Mellers, Barbara A. and Rohrbaugh, Nick and Chen, Eva},
	month = aug,
	year = {2014},
	pages = {290--295},
}

@article{roese_twenty_1993,
	title = {Twenty years of bogus pipeline research: {A} critical review and meta-analysis.},
	volume = {114},
	number = {2},
	journal = {Psychological Bulletin},
	author = {Roese, Neal J. and Jamieson, David W.},
	month = sep,
	year = {1993},
	pages = {363--375},
}

@book{osborne_introduction_2004,
	title = {An {Introduction} to {Game} {Theory}},
	publisher = {Oxford University Press},
	author = {Osborne, Martin J},
	year = {2004},
}

@article{krumpal_determinants_2013,
	title = {Determinants of social desirability bias in sensitive surveys: a literature review},
	volume = {47},
	copyright = {http://www.springer.com/tdm},
	issn = {0033-5177, 1573-7845},
	shorttitle = {Determinants of social desirability bias in sensitive surveys},
	url = {http://link.springer.com/10.1007/s11135-011-9640-9},
	doi = {10.1007/s11135-011-9640-9},
	language = {en},
	number = {4},
	urldate = {2024-09-05},
	journal = {Quality \& Quantity},
	author = {Krumpal, Ivar},
	month = jun,
	year = {2013},
	pages = {2025--2047},
}

@article{keren_facing_1987,
	title = {Facing uncertainty in the game of bridge: {A} calibration study},
	volume = {39},
	copyright = {https://www.elsevier.com/tdm/userlicense/1.0/},
	issn = {07495978},
	shorttitle = {Facing uncertainty in the game of bridge},
	url = {https://linkinghub.elsevier.com/retrieve/pii/0749597887900471},
	doi = {10.1016/0749-5978(87)90047-1},
	language = {en},
	number = {1},
	urldate = {2024-09-04},
	journal = {Organizational Behavior and Human Decision Processes},
	author = {Keren, Gideon},
	month = feb,
	year = {1987},
	pages = {98--114},
}

@article{johnson_defaults_2003,
	title = {Do {Defaults} {Save} {Lives}?},
	volume = {302},
	issn = {0036-8075, 1095-9203},
	url = {https://www.science.org/doi/10.1126/science.1091721},
	doi = {10.1126/science.1091721},
	language = {en},
	number = {5649},
	urldate = {2024-09-04},
	journal = {Science},
	author = {Johnson, Eric J. and Goldstein, Daniel},
	month = nov,
	year = {2003},
	pages = {1338--1339},
}

@article{baron_two_2014,
	title = {Two {Reasons} to {Make} {Aggregated} {Probability} {Forecasts} {More} {Extreme}},
	volume = {11},
	issn = {1545-8490, 1545-8504},
	url = {https://pubsonline.informs.org/doi/10.1287/deca.2014.0293},
	doi = {10.1287/deca.2014.0293},
	language = {en},
	number = {2},
	urldate = {2024-09-04},
	journal = {Decision Analysis},
	author = {Baron, Jonathan and Mellers, Barbara A. and Tetlock, Philip E. and Stone, Eric and Ungar, Lyle H.},
	month = jun,
	year = {2014},
	pages = {133--145},
}

@article{spann_sports_2009,
	title = {Sports forecasting: a comparison of the forecast accuracy of prediction markets, betting odds and tipsters},
	volume = {28},
	copyright = {http://onlinelibrary.wiley.com/termsAndConditions\#vor},
	issn = {0277-6693, 1099-131X},
	shorttitle = {Sports forecasting},
	url = {https://onlinelibrary.wiley.com/doi/10.1002/for.1091},
	doi = {10.1002/for.1091},
	language = {en},
	number = {1},
	urldate = {2024-09-04},
	journal = {Journal of Forecasting},
	author = {Spann, Martin and Skiera, Bernd},
	month = jan,
	year = {2009},
	pages = {55--72},
}

@article{kruger_quantifying_2024,
	title = {Quantifying subjective uncertainty in survey expectations},
	volume = {40},
	issn = {01692070},
	url = {https://linkinghub.elsevier.com/retrieve/pii/S0169207023000699},
	doi = {10.1016/j.ijforecast.2023.06.001},
	language = {en},
	number = {2},
	urldate = {2024-09-04},
	journal = {International Journal of Forecasting},
	author = {Krüger, Fabian and Pavlova, Lora},
	month = apr,
	year = {2024},
	pages = {796--810},
}

@article{estrella_how_2003,
	title = {How {Stable} is the {Predictive} {Power} of the {Yield} {Curve}? {Evidence} from {Germany} and the {United} {States}},
	volume = {85},
	issn = {0034-6535, 1530-9142},
	shorttitle = {How {Stable} is the {Predictive} {Power} of the {Yield} {Curve}?},
	url = {https://direct.mit.edu/rest/article/85/3/629-644/57434},
	doi = {10.1162/003465303322369777},
	language = {en},
	number = {3},
	urldate = {2024-09-04},
	journal = {Review of Economics and Statistics},
	author = {Estrella, Arturo and Rodrigues, Anthony P. and Schich, Sebastian},
	month = aug,
	year = {2003},
	pages = {629--644},
}

@article{dana_are_2019,
	title = {Are markets more accurate than polls? {The} surprising informational value of “just asking”},
	volume = {14},
	copyright = {http://creativecommons.org/licenses/by-nc-nd/3.0/},
	issn = {1930-2975},
	shorttitle = {Are markets more accurate than polls?},
	url = {https://www.cambridge.org/core/product/identifier/S1930297500003375/type/journal_article},
	doi = {10.1017/S1930297500003375},
	language = {en},
	number = {2},
	urldate = {2024-09-04},
	journal = {Judgment and Decision Making},
	author = {Dana, Jason and Atanasov, Pavel and Tetlock, Philip and Mellers, Barbara},
	month = mar,
	year = {2019},
	pages = {135--147},
}

@misc{feng_peer_2022,
	title = {Peer {Prediction} for {Learning} {Agents}},
	copyright = {Creative Commons Attribution 4.0 International},
	url = {https://arxiv.org/abs/2208.04433},
	doi = {10.48550/ARXIV.2208.04433},
	urldate = {2024-06-13},
	publisher = {arXiv},
	author = {Feng, Shi and Yu, Fang Yi and Chen, Yiling},
	year = {2022},
	note = {Version Number: 2},
}

@article{frey_motivation_2001,
	title = {Motivation {Crowding} {Theory}},
	volume = {15},
	issn = {0950-0804, 1467-6419},
	url = {https://onlinelibrary.wiley.com/doi/10.1111/1467-6419.00150},
	doi = {10.1111/1467-6419.00150},
	language = {en},
	number = {5},
	urldate = {2024-07-01},
	journal = {Journal of Economic Surveys},
	author = {Frey, Bruno S. and Jegen, Reto},
	month = dec,
	year = {2001},
	pages = {589--611},
}

@inproceedings{gao_trick_2014,
	address = {Palo Alto California USA},
	title = {Trick or treat: putting peer prediction to the test},
	isbn = {978-1-4503-2565-3},
	shorttitle = {Trick or treat},
	url = {https://dl.acm.org/doi/10.1145/2600057.2602865},
	doi = {10.1145/2600057.2602865},
	language = {en},
	urldate = {2024-06-13},
	booktitle = {Proceedings of the fifteenth {ACM} conference on {Economics} and computation},
	publisher = {ACM},
	author = {Gao, Alice and Mao, Andrew and Chen, Yiling and Adams, Ryan Prescott},
	month = jun,
	year = {2014},
	pages = {507--524},
}

@article{radas_whose_2019,
	title = {Whose data can we trust: {How} meta-predictions can be used to uncover credible respondents in survey data},
	volume = {14},
	issn = {1932-6203},
	shorttitle = {Whose data can we trust},
	url = {https://dx.plos.org/10.1371/journal.pone.0225432},
	doi = {10.1371/journal.pone.0225432},
	language = {en},
	number = {12},
	urldate = {2024-06-13},
	journal = {PLOS ONE},
	author = {Radas, Sonja and Prelec, Drazen},
	month = dec,
	year = {2019},
	pages = {e0225432},
}

@article{lee_testing_2018,
	title = {Testing the ability of the surprisingly popular method to predict {NFL} games},
	volume = {13},
	copyright = {http://creativecommons.org/licenses/3.0/},
	issn = {1930-2975},
	url = {https://www.cambridge.org/core/product/identifier/S1930297500009207/type/journal_article},
	doi = {10.1017/S1930297500009207},
	language = {en},
	number = {4},
	urldate = {2024-06-14},
	journal = {Judgment and Decision Making},
	author = {Lee, Michael D. and Danileiko, Irina and Vi, Julie},
	month = jul,
	year = {2018},
	pages = {322--333},
}

@article{kong_equilibrium_2018,
	title = {Equilibrium {Selection} in {Information} {Elicitation} without {Verification} via {Information} {Monotonicity}},
	volume = {94},
	copyright = {Creative Commons Attribution 3.0 Unported license, info:eu-repo/semantics/openAccess},
	issn = {1868-8969},
	url = {https://drops.dagstuhl.de/entities/document/10.4230/LIPIcs.ITCS.2018.13},
	doi = {10.4230/LIPICS.ITCS.2018.13},
	language = {en},
	urldate = {2024-06-13},
	journal = {LIPIcs, Volume 94, ITCS 2018},
	author = {Kong, Yuqing and Schoenebeck, Grant},
	collaborator = {Karlin, Anna R.},
	year = {2018},
	pages = {13:1--13:20},
}

@article{kong_more_2022,
	title = {More {Dominantly} {Truthful} {Multi}-{Task} {Peer} {Prediction} with a {Finite} {Number} of {Tasks}},
	volume = {215},
	copyright = {Creative Commons Attribution 4.0 International license, info:eu-repo/semantics/openAccess},
	issn = {1868-8969},
	url = {https://drops.dagstuhl.de/entities/document/10.4230/LIPIcs.ITCS.2022.95},
	doi = {10.4230/LIPICS.ITCS.2022.95},
	language = {en},
	urldate = {2024-06-13},
	journal = {LIPIcs, Volume 215, ITCS 2022},
	author = {Kong, Yuqing},
	collaborator = {Braverman, Mark},
	year = {2022},
	pages = {95:1--95:20},
}

@article{forsell_predicting_2019,
	title = {Predicting replication outcomes in the {Many} {Labs} 2 study},
	doi = {10.1016/j.joep.2018.10.009},
	journal = {Journal of Economic Psychology},
	author = {Forsell, Eskil and Viganola, Domenico and Pfeiffer, Thomas and Almenberg, Johan and Wilson, Brad and Chen, Yiling and Nosek, Brian A. and Johannesson, Magnus and Dreber, Anna},
	year = {2019},
}

@article{witkowski_incentive-compatible_2023,
	title = {Incentive-{Compatible} {Forecasting} {Competitions}},
	volume = {69},
	issn = {0025-1909, 1526-5501},
	url = {https://pubsonline.informs.org/doi/10.1287/mnsc.2022.4410},
	doi = {10.1287/mnsc.2022.4410},
	language = {en},
	number = {3},
	urldate = {2024-06-14},
	journal = {Management Science},
	author = {Witkowski, Jens and Freeman, Rupert and Vaughan, Jennifer Wortman and Pennock, David M. and Krause, Andreas},
	month = mar,
	year = {2023},
	pages = {1354--1374},
}

@article{satopaa_combining_2014,
	title = {Combining multiple probability predictions using a simple logit model},
	doi = {10.1016/j.ijforecast.2013.09.009},
	journal = {International Journal of Forecasting},
	author = {Satopää, Ville A. and Baron, Jonathan and Foster, Dean P. and Mellers, Barbara A. and Tetlock, Philip E. and Ungar, Lyle H.},
	year = {2014},
}

@inproceedings{kong_dominantly_2020,
	title = {Dominantly {Truthful} {Multi}-task {Peer} {Prediction} with a {Constant} {Number} of {Tasks}},
	isbn = {978-1-61197-599-4},
	booktitle = {Proceedings of the {Fourteenth} {Annual} {ACM}-{SIAM} {Symposium} on {Discrete} {Algorithms}},
	author = {Kong, Yuqing},
	year = {2020},
	pages = {2398 -- 2411},
}

@article{radanovic_robust_2013,
	title = {A robust {Bayesian} truth serum for non-binary signals},
	doi = {10.1609/aaai.v27i1.8677},
	journal = {AAAI Conference on Artificial Intelligence},
	author = {Radanovic, Goran and Faltings, Boi},
	year = {2013},
}

@article{witkowski_robust_2012,
	title = {A robust {Bayesian} truth serum for small populations},
	doi = {10.1609/aaai.v26i1.8261},
	journal = {AAAI Conference on Artificial Intelligence},
	author = {Witkowski, Jens and Parkes, David C.},
	year = {2012},
}

@article{frank_validating_2017,
	title = {Validating {Bayesian} truth serum in large-scale online human experiments.},
	doi = {10.1371/journal.pone.0177385},
	journal = {PLOS ONE},
	author = {Frank, Morgan R. and Cebrian, Manuel and Pickard, Galen and Rahwan, Iyad},
	year = {2017},
}

@article{gruetzemacher_forecasting_2021,
	title = {Forecasting {AI} progress: {A} research agenda},
	volume = {170},
	issn = {00401625},
	shorttitle = {Forecasting {AI} progress},
	url = {https://linkinghub.elsevier.com/retrieve/pii/S0040162521003413},
	doi = {10.1016/j.techfore.2021.120909},
	language = {en},
	urldate = {2024-06-14},
	journal = {Technological Forecasting and Social Change},
	author = {Gruetzemacher, Ross and Dorner, Florian E. and Bernaola-Alvarez, Niko and Giattino, Charlie and Manheim, David},
	month = sep,
	year = {2021},
	pages = {120909},
}

@article{mandal_effectiveness_2020,
	title = {The {Effectiveness} of {Peer} {Prediction} in {Long}-{Term} {Forecasting}},
	doi = {10.1609/aaai.v34i02.5591},
	journal = {AAAI Conference on Artificial Intelligence},
	author = {Mandal, Debmalya and Goran, Radanović and Parkes, David C.},
	year = {2020},
}

@article{clemen_combining_1989,
	title = {Combining forecasts: {A} review and annotated bibliography},
	doi = {10.1016/0169-2070(89)90012-5},
	journal = {International Journal of Forecasting},
	author = {Clemen, Robert T.},
	year = {1989},
}

@book{surowiecki_wisdom_2004,
	title = {The {Wisdom} of {Crowds}: {Why} the {Many} {Are} {Smarter} {Than} the {Few} and {How} {Collective} {Wisdom} {Shapes} {Business}, {Economies}, {Societies} and {Nations}},
	publisher = {Anchor},
	author = {Surowiecki, James},
	year = {2004},
}

@article{zhang_elicitability_2014,
	title = {Elicitability and knowledge-free elicitation with peer prediction},
	journal = {Adaptive Agents and Multi-Agent Systems},
	author = {Zhang, Peter and Chen, Yiling},
	year = {2014},
}

@article{zawojska_incentivizing_2022,
	title = {Incentivizing {Stated} {Preference} {Elicitation} with {Choice}-{Matching} in the {Field}},
	issn = {1556-5068},
	url = {https://www.ssrn.com/abstract=4052462},
	doi = {10.2139/ssrn.4052462},
	language = {en},
	urldate = {2024-06-14},
	journal = {SSRN Electronic Journal},
	author = {Zawojska, Ewa and Krawczyk, Michal Wiktor},
	year = {2022},
}

@article{weaver_creating_2013,
	title = {Creating {Truth}-{Telling} {Incentives} with the {Bayesian} {Truth} {Serum}},
	volume = {50},
	issn = {0022-2437, 1547-7193},
	url = {http://journals.sagepub.com/doi/10.1509/jmr.09.0039},
	doi = {10.1509/jmr.09.0039},
	language = {en},
	number = {3},
	urldate = {2024-06-14},
	journal = {Journal of Marketing Research},
	author = {Weaver, Ray and Prelec, Drazen},
	month = jun,
	year = {2013},
	pages = {289--302},
}

@article{cvitanic_honesty_2019,
	title = {Honesty via {Choice}-{Matching}},
	volume = {1},
	issn = {2640-205X, 2640-2068},
	url = {https://pubs.aeaweb.org/doi/10.1257/aeri.20180227},
	doi = {10.1257/aeri.20180227},
	language = {en},
	number = {2},
	urldate = {2024-06-14},
	journal = {American Economic Review: Insights},
	author = {Cvitanić, Jakša and Prelec, Dražen and Riley, Blake and Tereick, Benjamin},
	month = sep,
	year = {2019},
	pages = {179--192},
}

@inproceedings{shaw_designing_2011,
	address = {Hangzhou China},
	title = {Designing incentives for inexpert human raters},
	isbn = {978-1-4503-0556-3},
	url = {https://dl.acm.org/doi/10.1145/1958824.1958865},
	doi = {10.1145/1958824.1958865},
	language = {en},
	urldate = {2024-06-14},
	booktitle = {Proceedings of the {ACM} 2011 conference on {Computer} supported cooperative work},
	publisher = {ACM},
	author = {Shaw, Aaron D. and Horton, John J. and Chen, Daniel L.},
	month = mar,
	year = {2011},
	pages = {275--284},
}

@article{zhou_projected_2017,
	title = {Projected prevalence of car-sharing in four {Asian}-{Pacific} countries in 2030: {What} the experts think},
	volume = {84},
	issn = {0968090X},
	shorttitle = {Projected prevalence of car-sharing in four {Asian}-{Pacific} countries in 2030},
	url = {https://linkinghub.elsevier.com/retrieve/pii/S0968090X17302334},
	doi = {10.1016/j.trc.2017.08.023},
	language = {en},
	urldate = {2024-06-14},
	journal = {Transportation Research Part C: Emerging Technologies},
	author = {Zhou, Fan and Zheng, Zuduo and Whitehead, Jake and Perrons, Robert and Page, Lionel and Washington, Simon},
	month = nov,
	year = {2017},
	pages = {158--177},
}

@misc{srinivasan_auctions_2021,
	title = {Auctions and {Peer} {Prediction} for {Academic} {Peer} {Review}},
	copyright = {Creative Commons Attribution 4.0 International},
	url = {https://arxiv.org/abs/2109.00923},
	doi = {10.48550/ARXIV.2109.00923},
	urldate = {2024-06-14},
	publisher = {arXiv},
	author = {Srinivasan, Siddarth and Morgenstern, Jamie},
	year = {2021},
	note = {Version Number: 2},
}

@article{baillon_incentives_2022,
	title = {Incentives in surveys},
	volume = {93},
	issn = {01674870},
	url = {https://linkinghub.elsevier.com/retrieve/pii/S0167487022000642},
	doi = {10.1016/j.joep.2022.102552},
	language = {en},
	urldate = {2024-06-14},
	journal = {Journal of Economic Psychology},
	author = {Baillon, Aurélien and Bleichrodt, Han and Granic, Georg D.},
	month = dec,
	year = {2022},
	pages = {102552},
}

@article{schoenegger_taking_2022,
	title = {Taking a {Closer} {Look} at the {Bayesian} {Truth} {Serum}: {A} {Registered} {Report}},
	volume = {69},
	issn = {1618-3169, 2190-5142},
	shorttitle = {Taking a {Closer} {Look} at the {Bayesian} {Truth} {Serum}},
	url = {https://econtent.hogrefe.com/doi/10.1027/1618-3169/a000558},
	doi = {10.1027/1618-3169/a000558},
	language = {en},
	number = {4},
	urldate = {2024-06-14},
	journal = {Experimental Psychology},
	author = {Schoenegger, Philipp and Verheyen, Steven},
	month = jul,
	year = {2022},
	pages = {226--239},
}

@article{van_de_schoot_use_2021,
	title = {The {Use} of {Questionable} {Research} {Practices} to {Survive} in {Academia} {Examined} {With} {Expert} {Elicitation}, {Prior}-{Data} {Conflicts}, {Bayes} {Factors} for {Replication} {Effects}, and the {Bayes} {Truth} {Serum}},
	volume = {12},
	issn = {1664-1078},
	url = {https://www.frontiersin.org/articles/10.3389/fpsyg.2021.621547/full},
	doi = {10.3389/fpsyg.2021.621547},
	urldate = {2024-06-14},
	journal = {Frontiers in Psychology},
	author = {Van De Schoot, Rens and Winter, Sonja D. and Griffioen, Elian and Grimmelikhuijsen, Stephan and Arts, Ingrid and Veen, Duco and Grandfield, Elizabeth M. and Tummers, Lars G.},
	month = nov,
	year = {2021},
	pages = {621547},
}

@article{schoenegger_experimental_2023,
	title = {Experimental {Philosophy} and the {Incentivisation} {Challenge}: a {Proposed} {Application} of the {Bayesian} {Truth} {Serum}},
	volume = {14},
	issn = {1878-5158, 1878-5166},
	shorttitle = {Experimental {Philosophy} and the {Incentivisation} {Challenge}},
	url = {https://link.springer.com/10.1007/s13164-021-00571-4},
	doi = {10.1007/s13164-021-00571-4},
	language = {en},
	number = {1},
	urldate = {2024-06-14},
	journal = {Review of Philosophy and Psychology},
	author = {Schoenegger, Philipp},
	month = mar,
	year = {2023},
	pages = {295--320},
}

@article{loughran_incentivizing_2014,
	title = {Incentivizing {Responses} to {Self}-report {Questions} in {Perceptual} {Deterrence} {Studies}: {An} {Investigation} of the {Validity} of {Deterrence} {Theory} {Using} {Bayesian} {Truth} {Serum}},
	volume = {30},
	copyright = {http://www.springer.com/tdm},
	issn = {0748-4518, 1573-7799},
	shorttitle = {Incentivizing {Responses} to {Self}-report {Questions} in {Perceptual} {Deterrence} {Studies}},
	url = {http://link.springer.com/10.1007/s10940-014-9219-4},
	doi = {10.1007/s10940-014-9219-4},
	language = {en},
	number = {4},
	urldate = {2024-06-14},
	journal = {Journal of Quantitative Criminology},
	author = {Loughran, Thomas A. and Paternoster, Ray and Thomas, Kyle J.},
	month = dec,
	year = {2014},
	pages = {677--707},
}

@article{zhou_long-term_2019,
	title = {Long-term forecasts for energy commodities price: {What} the experts think},
	volume = {84},
	issn = {01409883},
	shorttitle = {Long-term forecasts for energy commodities price},
	url = {https://linkinghub.elsevier.com/retrieve/pii/S0140988319302658},
	doi = {10.1016/j.eneco.2019.104484},
	language = {en},
	urldate = {2024-06-14},
	journal = {Energy Economics},
	author = {Zhou, Fan and Page, Lionel and Perrons, Robert K. and Zheng, Zuduo and Washington, Simon},
	month = oct,
	year = {2019},
	pages = {104484},
}

@article{john_measuring_2012,
	title = {Measuring the {Prevalence} of {Questionable} {Research} {Practices} {With} {Incentives} for {Truth} {Telling}},
	volume = {23},
	issn = {0956-7976, 1467-9280},
	url = {http://journals.sagepub.com/doi/10.1177/0956797611430953},
	doi = {10.1177/0956797611430953},
	language = {en},
	number = {5},
	urldate = {2024-06-14},
	journal = {Psychological Science},
	author = {John, Leslie K. and Loewenstein, George and Prelec, Drazen},
	month = may,
	year = {2012},
	pages = {524--532},
}

@article{prelec_bayesian_2004,
	title = {A {Bayesian} {Truth} {Serum} for {Subjective} {Data}},
	doi = {10.1126/science.1102081},
	journal = {Science},
	author = {Prelec, Dražen},
	year = {2004},
}

@article{papakonstantinou_mechanism_2011,
	title = {Mechanism design for the truthful elicitation of costly probabilistic estimates in distributed information systems},
	volume = {175},
	copyright = {https://www.elsevier.com/tdm/userlicense/1.0/},
	issn = {00043702},
	url = {https://linkinghub.elsevier.com/retrieve/pii/S0004370210001773},
	doi = {10.1016/j.artint.2010.10.007},
	language = {en},
	number = {2},
	urldate = {2024-06-13},
	journal = {Artificial Intelligence},
	author = {Papakonstantinou, Athanasios and Rogers, Alex and Gerding, Enrico H. and Jennings, Nicholas R.},
	month = feb,
	year = {2011},
	pages = {648--672},
}

@inproceedings{ghosh_buying_2014,
	address = {Palo Alto California USA},
	title = {Buying private data without verification},
	isbn = {978-1-4503-2565-3},
	url = {https://dl.acm.org/doi/10.1145/2600057.2602902},
	doi = {10.1145/2600057.2602902},
	language = {en},
	urldate = {2024-06-13},
	booktitle = {Proceedings of the fifteenth {ACM} conference on {Economics} and computation},
	publisher = {ACM},
	author = {Ghosh, Arpita and Ligett, Katrina and Roth, Aaron and Schoenebeck, Grant},
	month = jun,
	year = {2014},
	pages = {931--948},
}

@inproceedings{jurca_incentives_2008,
	address = {Chicago Il USA},
	title = {Incentives for expressing opinions in online polls},
	isbn = {978-1-60558-169-9},
	url = {https://dl.acm.org/doi/10.1145/1386790.1386812},
	doi = {10.1145/1386790.1386812},
	language = {en},
	urldate = {2024-06-13},
	booktitle = {Proceedings of the 9th {ACM} conference on {Electronic} commerce},
	publisher = {ACM},
	author = {Jurca, Radu and Faltings, Boi},
	month = jul,
	year = {2008},
	pages = {119--128},
}

@article{kong_dominantly_2024,
	title = {Dominantly {Truthful} {Peer} {Prediction} {Mechanisms} with a {Finite} {Number} of {Tasks}},
	volume = {71},
	issn = {0004-5411, 1557-735X},
	url = {https://dl.acm.org/doi/10.1145/3638239},
	doi = {10.1145/3638239},
	language = {en},
	number = {2},
	urldate = {2024-06-13},
	journal = {Journal of the ACM},
	author = {Kong, Yuqing},
	month = apr,
	year = {2024},
	pages = {1--49},
}

@article{kong_information_2019,
	title = {An {Information} {Theoretic} {Framework} {For} {Designing} {Information} {Elicitation} {Mechanisms} {That} {Reward} {Truth}-telling},
	volume = {7},
	issn = {2167-8375, 2167-8383},
	url = {https://dl.acm.org/doi/10.1145/3296670},
	doi = {10.1145/3296670},
	language = {en},
	number = {1},
	urldate = {2024-06-13},
	journal = {ACM Transactions on Economics and Computation},
	author = {Kong, Yuqing and Schoenebeck, Grant},
	month = feb,
	year = {2019},
	pages = {1--33},
}

@inproceedings{agarwal_peer_2017,
	address = {Cambridge Massachusetts USA},
	title = {Peer {Prediction} with {Heterogeneous} {Users}},
	isbn = {978-1-4503-4527-9},
	url = {https://dl.acm.org/doi/10.1145/3033274.3085127},
	doi = {10.1145/3033274.3085127},
	language = {en},
	urldate = {2024-06-13},
	booktitle = {Proceedings of the 2017 {ACM} {Conference} on {Economics} and {Computation}},
	publisher = {ACM},
	author = {Agarwal, Arpit and Mandal, Debmalya and Parkes, David C. and Shah, Nisarg},
	month = jun,
	year = {2017},
	pages = {81--98},
}

@inproceedings{witkowski_learning_2013,
	title = {Learning the {Prior} in {Minimal} {Peer} {Prediction}},
	language = {en},
	booktitle = {Proceedings of the 3rd {Workshop} on {Social} {Computing} and {User} {Generated} {Content} at the {ACM} {Conference} on {Electronic} {Commerce}},
	author = {Witkowski, Jens and Parkes, David C},
	year = {2013},
}

@inproceedings{gao_incentivizing_2020,
	address = {Yokohama, Japan},
	title = {Incentivizing {Evaluation} with {Peer} {Prediction} and {Limited} {Access} to {Ground} {Truth}},
	isbn = {978-0-9992411-6-5},
	url = {https://www.ijcai.org/proceedings/2020/723},
	doi = {10.24963/ijcai.2020/723},
	language = {en},
	urldate = {2024-06-13},
	booktitle = {Proceedings of the {Twenty}-{Ninth} {International} {Joint} {Conference} on {Artificial} {Intelligence}},
	publisher = {International Joint Conferences on Artificial Intelligence Organization},
	author = {Gao, Alice and Wright, James and Leyton-Brown, Kevin},
	month = jul,
	year = {2020},
	pages = {5140--5144},
}

@article{liu_sequential_2017,
	title = {Sequential {Peer} {Prediction}: {Learning} to {Elicit} {Effort} using {Posted} {Prices}},
	volume = {31},
	issn = {2374-3468, 2159-5399},
	shorttitle = {Sequential {Peer} {Prediction}},
	url = {https://ojs.aaai.org/index.php/AAAI/article/view/10619},
	doi = {10.1609/aaai.v31i1.10619},
	number = {1},
	urldate = {2024-06-13},
	journal = {Proceedings of the AAAI Conference on Artificial Intelligence},
	author = {Liu, Yang and Chen, Yiling},
	month = feb,
	year = {2017},
}

@inproceedings{goel_personalized_2020,
	title = {Personalized {Peer} {Truth} {Serum} for {Eliciting} {Multi}-{Attribute} {Personal} {Data}},
	url = {http://proceedings.mlr.press/v115/goel20a.html},
	language = {en},
	booktitle = {Proceedings of {The} 35th {Uncertainty} in {Artificial} {Intelligence} {Conference}},
	author = {Goel, Naman and Faltings, Boi},
	year = {2020},
}

@article{radanovic_incentives_2015,
	title = {Incentives for {Subjective} {Evaluations} with {Private} {Beliefs}},
	volume = {29},
	issn = {2374-3468, 2159-5399},
	url = {https://ojs.aaai.org/index.php/AAAI/article/view/9311},
	doi = {10.1609/aaai.v29i1.9311},
	number = {1},
	urldate = {2024-06-13},
	journal = {Proceedings of the AAAI Conference on Artificial Intelligence},
	author = {Radanovic, Goran and Faltings, Boi},
	month = feb,
	year = {2015},
}

@article{radanovic_incentives_2016,
	title = {Incentives for {Effort} in {Crowdsourcing} {Using} the {Peer} {Truth} {Serum}},
	volume = {7},
	issn = {2157-6904, 2157-6912},
	url = {https://dl.acm.org/doi/10.1145/2856102},
	doi = {10.1145/2856102},
	language = {en},
	number = {4},
	urldate = {2024-06-13},
	journal = {ACM Transactions on Intelligent Systems and Technology},
	author = {Radanovic, Goran and Faltings, Boi and Jurca, Radu},
	month = jul,
	year = {2016},
	pages = {1--28},
}

@inproceedings{dasgupta_crowdsourced_2013,
	address = {Rio de Janeiro Brazil},
	title = {Crowdsourced judgement elicitation with endogenous proficiency},
	isbn = {978-1-4503-2035-1},
	url = {https://dl.acm.org/doi/10.1145/2488388.2488417},
	doi = {10.1145/2488388.2488417},
	language = {en},
	urldate = {2024-06-13},
	booktitle = {Proceedings of the 22nd international conference on {World} {Wide} {Web}},
	publisher = {ACM},
	author = {Dasgupta, Anirban and Ghosh, Arpita},
	month = may,
	year = {2013},
	pages = {319--330},
}

@article{miller_eliciting_2005,
	title = {Eliciting {Informative} {Feedback}: {The} {Peer}-{Prediction} {Method}},
	volume = {51},
	issn = {0025-1909, 1526-5501},
	shorttitle = {Eliciting {Informative} {Feedback}},
	url = {https://pubsonline.informs.org/doi/10.1287/mnsc.1050.0379},
	doi = {10.1287/mnsc.1050.0379},
	language = {en},
	number = {9},
	urldate = {2024-06-13},
	journal = {Management Science},
	author = {Miller, Nolan and Resnick, Paul and Zeckhauser, Richard},
	month = sep,
	year = {2005},
	pages = {1359--1373},
}

@inproceedings{shnayder_informed_2016,
	address = {Maastricht The Netherlands},
	title = {Informed {Truthfulness} in {Multi}-{Task} {Peer} {Prediction}},
	isbn = {978-1-4503-3936-0},
	url = {https://dl.acm.org/doi/10.1145/2940716.2940790},
	doi = {10.1145/2940716.2940790},
	language = {en},
	urldate = {2024-06-13},
	booktitle = {Proceedings of the 2016 {ACM} {Conference} on {Economics} and {Computation}},
	publisher = {ACM},
	author = {Shnayder, Victor and Agarwal, Arpit and Frongillo, Rafael and Parkes, David C.},
	month = jul,
	year = {2016},
	pages = {179--196},
}

@inproceedings{witkowski_peer_2012,
	address = {Valencia Spain},
	title = {Peer prediction without a common prior},
	isbn = {978-1-4503-1415-2},
	url = {https://dl.acm.org/doi/10.1145/2229012.2229085},
	doi = {10.1145/2229012.2229085},
	language = {en},
	urldate = {2024-06-13},
	booktitle = {Proceedings of the 13th {ACM} {Conference} on {Electronic} {Commerce}},
	publisher = {ACM},
	author = {Witkowski, Jens and Parkes, David C.},
	month = jun,
	year = {2012},
	pages = {964--981},
}

@article{radanovic_incentives_2014,
	title = {Incentives for {Truthful} {Information} {Elicitation} of {Continuous} {Signals}},
	volume = {28},
	issn = {2374-3468, 2159-5399},
	url = {https://ojs.aaai.org/index.php/AAAI/article/view/8797},
	doi = {10.1609/aaai.v28i1.8797},
	number = {1},
	urldate = {2024-06-13},
	journal = {Proceedings of the AAAI Conference on Artificial Intelligence},
	author = {Radanovic, Goran and Faltings, Boi},
	month = jun,
	year = {2014},
}

@article{jurca_mechanisms_2009,
	title = {Mechanisms for {Making} {Crowds} {Truthful}},
	volume = {34},
	issn = {1076-9757},
	url = {https://jair.org/index.php/jair/article/view/10590},
	doi = {10.1613/jair.2621},
	urldate = {2024-06-13},
	journal = {Journal of Artificial Intelligence Research},
	author = {Jurca, R. and Faltings, B.},
	month = mar,
	year = {2009},
	pages = {209--253},
}

@article{wang_forecast_2023,
	title = {Forecast combinations: {An} over 50-year review},
	volume = {39},
	issn = {01692070},
	shorttitle = {Forecast combinations},
	url = {https://linkinghub.elsevier.com/retrieve/pii/S0169207022001480},
	doi = {10.1016/j.ijforecast.2022.11.005},
	language = {en},
	number = {4},
	urldate = {2024-06-13},
	journal = {International Journal of Forecasting},
	author = {Wang, Xiaoqian and Hyndman, Rob J. and Li, Feng and Kang, Yanfei},
	month = oct,
	year = {2023},
	pages = {1518--1547},
}

@article{charness_experimental_2021,
	title = {Experimental methods: {Eliciting} beliefs},
	volume = {189},
	issn = {01672681},
	shorttitle = {Experimental methods},
	url = {https://linkinghub.elsevier.com/retrieve/pii/S0167268121002717},
	doi = {10.1016/j.jebo.2021.06.032},
	language = {en},
	urldate = {2024-06-13},
	journal = {Journal of Economic Behavior \& Organization},
	author = {Charness, Gary and Gneezy, Uri and Rasocha, Vlastimil},
	month = sep,
	year = {2021},
	pages = {234--256},
}

@article{liu_surrogate_2022,
	title = {Surrogate {Scoring} {Rules}},
	volume = {10},
	issn = {2167-8375, 2167-8383},
	url = {https://dl.acm.org/doi/10.1145/3565559},
	doi = {10.1145/3565559},
	language = {en},
	number = {3},
	urldate = {2024-06-13},
	journal = {ACM Transactions on Economics and Computation},
	author = {Liu, Yang and Wang, Juntao and Chen, Yiling},
	month = sep,
	year = {2022},
	pages = {1--36},
}

@inproceedings{schoenebeck_information_2021,
	address = {Ljubljana Slovenia},
	title = {Information {Elicitation} from {Rowdy} {Crowds}},
	isbn = {978-1-4503-8312-7},
	url = {https://dl.acm.org/doi/10.1145/3442381.3449840},
	doi = {10.1145/3442381.3449840},
	language = {en},
	urldate = {2024-06-13},
	booktitle = {Proceedings of the {Web} {Conference} 2021},
	publisher = {ACM},
	author = {Schoenebeck, Grant and Yu, Fang-Yi and Zhang, Yichi},
	month = apr,
	year = {2021},
	pages = {3974--3986},
}

@article{rutchick_does_2020,
	title = {Does the “surprisingly popular” method yield accurate crowdsourced predictions?},
	volume = {5},
	issn = {2365-7464},
	url = {https://cognitiveresearchjournal.springeropen.com/articles/10.1186/s41235-020-00256-z},
	doi = {10.1186/s41235-020-00256-z},
	language = {en},
	number = {1},
	urldate = {2024-06-13},
	journal = {Cognitive Research: Principles and Implications},
	author = {Rutchick, Abraham M. and Ross, Bryan J. and Calvillo, Dustin P. and Mesick, Catherine C.},
	month = dec,
	year = {2020},
	pages = {57},
}

@article{wilkening_hidden_2022,
	title = {Hidden {Experts} in the {Crowd}: {Using} {Meta}-{Predictions} to {Leverage} {Expertise} in {Single}-{Question} {Prediction} {Problems}},
	volume = {68},
	issn = {0025-1909, 1526-5501},
	shorttitle = {Hidden {Experts} in the {Crowd}},
	url = {https://pubsonline.informs.org/doi/10.1287/mnsc.2020.3919},
	doi = {10.1287/mnsc.2020.3919},
	language = {en},
	number = {1},
	urldate = {2024-06-13},
	journal = {Management Science},
	author = {Wilkening, Tom and Martinie, Marcellin and Howe, Piers D. L.},
	month = jan,
	year = {2022},
	pages = {487--508},
}

@article{atanasov_full_2023,
	title = {Full {Accuracy} {Scoring} {Accelerates} the {Discovery} of {Skilled} {Forecasters}},
	issn = {1556-5068},
	url = {https://www.ssrn.com/abstract=4357367},
	doi = {10.2139/ssrn.4357367},
	language = {en},
	urldate = {2024-06-13},
	journal = {SSRN Electronic Journal},
	author = {Atanasov, Pavel D. and Karger, Ezra and Tetlock, Philip},
	year = {2023},
}

@article{martinie_using_2020,
	title = {Using meta-predictions to identify experts in the crowd when past performance is unknown},
	volume = {15},
	issn = {1932-6203},
	url = {https://dx.plos.org/10.1371/journal.pone.0232058},
	doi = {10.1371/journal.pone.0232058},
	language = {en},
	number = {4},
	urldate = {2024-06-13},
	journal = {PLOS ONE},
	author = {Martinie, Marcellin and Wilkening, Tom and Howe, Piers D. L.},
	editor = {Schwenker, Friedhelm},
	month = apr,
	year = {2020},
	pages = {e0232058},
}

@article{prelec_solution_2017,
	title = {A solution to the single-question crowd wisdom problem},
	volume = {541},
	issn = {0028-0836, 1476-4687},
	url = {https://www.nature.com/articles/nature21054},
	doi = {10.1038/nature21054},
	language = {en},
	number = {7638},
	urldate = {2024-06-13},
	journal = {Nature},
	author = {Prelec, Dražen and Seung, H. Sebastian and McCoy, John},
	month = jan,
	year = {2017},
	pages = {532--535},
}

@article{palley_extracting_2019,
	title = {Extracting the {Wisdom} of {Crowds} {When} {Information} {Is} {Shared}},
	issn = {0025-1909, 1526-5501},
	url = {http://pubsonline.informs.org/doi/10.1287/mnsc.2018.3047},
	doi = {10.1287/mnsc.2018.3047},
	language = {en},
	urldate = {2024-06-13},
	journal = {Management Science},
	author = {Palley, Asa B. and Soll, Jack B.},
	month = feb,
	year = {2019},
	pages = {mnsc.2018.3047},
}

@article{palley_boosting_2023,
	title = {Boosting the {Wisdom} of {Crowds} {Within} a {Single} {Judgment} {Problem}: {Weighted} {Averaging} {Based} on {Peer} {Predictions}},
	volume = {69},
	issn = {0025-1909, 1526-5501},
	shorttitle = {Boosting the {Wisdom} of {Crowds} {Within} a {Single} {Judgment} {Problem}},
	url = {https://pubsonline.informs.org/doi/10.1287/mnsc.2022.4648},
	doi = {10.1287/mnsc.2022.4648},
	language = {en},
	number = {9},
	urldate = {2024-06-13},
	journal = {Management Science},
	author = {Palley, Asa B. and Satopää, Ville A.},
	month = sep,
	year = {2023},
	pages = {5128--5146},
}

@article{wang_forecast_2021,
	title = {Forecast {Aggregation} via {Peer} {Prediction}},
	volume = {9},
	issn = {2769-1349, 2769-1330},
	url = {https://ojs.aaai.org/index.php/HCOMP/article/view/18946},
	doi = {10.1609/hcomp.v9i1.18946},
	urldate = {2024-06-13},
	journal = {Proceedings of the AAAI Conference on Human Computation and Crowdsourcing},
	author = {Wang, Juntao and Liu, Yang and Chen, Yiling},
	month = oct,
	year = {2021},
	pages = {131--142},
}

@article{waggoner_output_2014,
	title = {Output {Agreement} {Mechanisms} and {Common} {Knowledge}},
	volume = {2},
	issn = {2769-1349, 2769-1330},
	url = {https://ojs.aaai.org/index.php/HCOMP/article/view/13151},
	doi = {10.1609/hcomp.v2i1.13151},
	urldate = {2024-06-11},
	journal = {Proceedings of the AAAI Conference on Human Computation and Crowdsourcing},
	author = {Waggoner, Bo and Chen, Yiling},
	month = sep,
	year = {2014},
	pages = {220--226},
}

@article{court_two_2018,
	title = {Two information aggregation mechanisms for predicting the opening weekend box office revenues of films: {Boxoffice} {Prophecy} and {Guess} of {Guesses}},
	volume = {65},
	issn = {0938-2259, 1432-0479},
	shorttitle = {Two information aggregation mechanisms for predicting the opening weekend box office revenues of films},
	url = {http://link.springer.com/10.1007/s00199-017-1036-1},
	doi = {10.1007/s00199-017-1036-1},
	language = {en},
	number = {1},
	urldate = {2024-06-11},
	journal = {Economic Theory},
	author = {Court, David and Gillen, Benjamin and McKenzie, Jordi and Plott, Charles R.},
	month = jan,
	year = {2018},
	pages = {25--54},
}

@article{carvalho_inducing_2017,
	title = {Inducing honest reporting of private information in the presence of social projection.},
	volume = {4},
	issn = {2325-9973, 2325-9965},
	url = {https://doi.apa.org/doi/10.1037/dec0000052},
	doi = {10.1037/dec0000052},
	language = {en},
	number = {1},
	urldate = {2024-06-11},
	journal = {Decision},
	author = {Carvalho, Arthur and Dimitrov, Stanko and Larson, Kate},
	month = jan,
	year = {2017},
	pages = {25--51},
}

@article{witkowski_proper_2017,
	title = {Proper {Proxy} {Scoring} {Rules}},
	volume = {31},
	issn = {2374-3468, 2159-5399},
	url = {https://ojs.aaai.org/index.php/AAAI/article/view/10590},
	doi = {10.1609/aaai.v31i1.10590},
	number = {1},
	urldate = {2024-06-11},
	journal = {Proceedings of the AAAI Conference on Artificial Intelligence},
	author = {Witkowski, Jens and Atanasov, Pavel and Ungar, Lyle and Krause, Andreas},
	month = feb,
	year = {2017},
}

@article{karger_reciprocal_2021,
	title = {Reciprocal {Scoring}: {A} {Method} for {Forecasting} {Unanswerable} {Questions}},
	issn = {1556-5068},
	shorttitle = {Reciprocal {Scoring}},
	url = {https://www.ssrn.com/abstract=3954498},
	doi = {10.2139/ssrn.3954498},
	language = {en},
	urldate = {2024-06-11},
	journal = {SSRN Electronic Journal},
	author = {Karger, Ezra and Monrad, Joshua and Mellers, Barb and Tetlock, Philip},
	year = {2021},
}

@inproceedings{huang_enhancing_2013,
	address = {San Antonio Texas USA},
	title = {Enhancing reliability using peer consistency evaluation in human computation},
	isbn = {978-1-4503-1331-5},
	url = {https://dl.acm.org/doi/10.1145/2441776.2441847},
	doi = {10.1145/2441776.2441847},
	language = {en},
	urldate = {2024-06-11},
	booktitle = {Proceedings of the 2013 conference on {Computer} supported cooperative work},
	publisher = {ACM},
	author = {Huang, Shih-Wen and Fu, Wai-Tat},
	month = feb,
	year = {2013},
	pages = {639--648},
}

@article{von_ahn_designing_2008,
	title = {Designing games with a purpose},
	volume = {51},
	issn = {0001-0782, 1557-7317},
	url = {https://dl.acm.org/doi/10.1145/1378704.1378719},
	doi = {10.1145/1378704.1378719},
	language = {en},
	number = {8},
	urldate = {2024-06-11},
	journal = {Communications of the ACM},
	author = {Von Ahn, Luis and Dabbish, Laura},
	month = aug,
	year = {2008},
	pages = {58--67},
}

@article{kamble_square_2023,
	title = {The {Square} {Root} {Agreement} {Rule} for {Incentivizing} {Truthful} {Feedback} on {Online} {Platforms}},
	volume = {69},
	issn = {0025-1909, 1526-5501},
	url = {https://pubsonline.informs.org/doi/10.1287/mnsc.2022.4375},
	doi = {10.1287/mnsc.2022.4375},
	language = {en},
	number = {1},
	urldate = {2024-06-11},
	journal = {Management Science},
	author = {Kamble, Vijay and Shah, Nihar and Marn, David and Parekh, Abhay and Ramchandran, Kannan},
	month = jan,
	year = {2023},
	pages = {377--403},
}

@inproceedings{von_ahn_labeling_2004,
	address = {Vienna Austria},
	title = {Labeling images with a computer game},
	isbn = {978-1-58113-702-6},
	url = {https://dl.acm.org/doi/10.1145/985692.985733},
	doi = {10.1145/985692.985733},
	language = {en},
	urldate = {2024-06-11},
	booktitle = {Proceedings of the {SIGCHI} {Conference} on {Human} {Factors} in {Computing} {Systems}},
	publisher = {ACM},
	author = {Von Ahn, Luis and Dabbish, Laura},
	month = apr,
	year = {2004},
	pages = {319--326},
}

@article{baillon_simple_2021,
	title = {Simple bets to elicit private signals},
	volume = {16},
	copyright = {http://creativecommons.org/licenses/by-nc/4.0/},
	issn = {1933-6837},
	url = {https://econtheory.org/ojs/index.php/te/article/view/4343},
	doi = {10.3982/TE4343},
	language = {en},
	number = {3},
	urldate = {2024-06-11},
	journal = {Theoretical Economics},
	author = {Baillon, Aurélien and Xu, Yan},
	year = {2021},
	pages = {777--797},
}

@article{slamka_secondgeneration_2012,
	title = {Second‐{Generation} {Prediction} {Markets} for {Information} {Aggregation}: {A} {Comparison} of {Payoff} {Mechanisms}},
	volume = {31},
	copyright = {http://onlinelibrary.wiley.com/termsAndConditions\#vor},
	issn = {0277-6693, 1099-131X},
	shorttitle = {Second‐{Generation} {Prediction} {Markets} for {Information} {Aggregation}},
	url = {https://onlinelibrary.wiley.com/doi/10.1002/for.1225},
	doi = {10.1002/for.1225},
	language = {en},
	number = {6},
	urldate = {2024-06-11},
	journal = {Journal of Forecasting},
	author = {Slamka, Christian and Jank, Wolfgang and Skiera, Bernd},
	month = sep,
	year = {2012},
	pages = {469--489},
}

@article{baillon_bayesian_2017,
	title = {Bayesian markets to elicit private information},
	volume = {114},
	issn = {0027-8424, 1091-6490},
	url = {https://pnas.org/doi/full/10.1073/pnas.1703486114},
	doi = {10.1073/pnas.1703486114},
	language = {en},
	number = {30},
	urldate = {2024-06-11},
	journal = {Proceedings of the National Academy of Sciences},
	author = {Baillon, Aurélien},
	month = jul,
	year = {2017},
	pages = {7958--7962},
}

@misc{srinivasan_self-resolving_2023,
	title = {Self-{Resolving} {Prediction} {Markets} for {Unverifiable} {Outcomes}},
	copyright = {Creative Commons Attribution 4.0 International},
	url = {https://arxiv.org/abs/2306.04305},
	doi = {10.48550/ARXIV.2306.04305},
	urldate = {2024-06-11},
	publisher = {arXiv},
	author = {Srinivasan, Siddarth and Karger, Ezra and Chen, Yiling},
	year = {2023},
	note = {Version Number: 1},
}

@article{gneiting_strictly_2007,
	title = {Strictly {Proper} {Scoring} {Rules}, {Prediction}, and {Estimation}},
	volume = {102},
	issn = {0162-1459, 1537-274X},
	url = {http://www.tandfonline.com/doi/abs/10.1198/016214506000001437},
	doi = {10.1198/016214506000001437},
	language = {en},
	number = {477},
	urldate = {2024-05-24},
	journal = {Journal of the American Statistical Association},
	author = {Gneiting, Tilmann and Raftery, Adrian E},
	month = mar,
	year = {2007},
	pages = {359--378},
}

@article{gneiting_probabilistic_2014,
	title = {Probabilistic {Forecasting}},
	volume = {1},
	issn = {2326-8298, 2326-831X},
	url = {https://www.annualreviews.org/doi/10.1146/annurev-statistics-062713-085831},
	doi = {10.1146/annurev-statistics-062713-085831},
	language = {en},
	number = {1},
	urldate = {2024-05-24},
	journal = {Annual Review of Statistics and Its Application},
	author = {Gneiting, Tilmann and Katzfuss, Matthias},
	month = jan,
	year = {2014},
	pages = {125--151},
}

\newpage

\section*{Appendix A}\label{sec: Appendix}

\subsection*{Tabloid overview of mechanisms}

The following tables contain most mechanisms that are mentioned in some form in the main text. The mechanisms are assigned a \ding{51} in the category 'Nash' in case that truthfulness is a BNE of the mechanisms and in the category 'Emp.' in case that empirical evidence regarding the mechanism exists. The \ding{51} is in brackets for the Peer-Prediction because the only existing empirical evidence brings forward negative results.

\newpage

\begin{landscape}
    
\begin{table}
    \centering
    \caption{Peer-Prediction Mechanisms \& Truth serums}
    \rowcolors{2}{}{gray!10}
    \begin{tabular}{*5c}
        \toprule
  
        Reference  & Mechanism name & Emp. & Nash  \\
        \midrule
        \cite{prelec_bayesian_2004} & Bayesian Truth Serum & \ding{51} & \ding{51}  \\
        \cite{miller_eliciting_2005} & Peer-Prediction & (\ding{51}) & \ding{51} \\
        \cite{jurca_incentives_2008} & - &  & \ding{51}  \\
        \cite{papakonstantinou_mechanism_2011} & Fusion &  & \ding{51}\\
        \cite{witkowski_robust_2012} & Robust BTS &  & \ding{51}  \\
        \cite{dasgupta_crowdsourced_2013} & - &  & \ding{51} \\
        \cite{ghosh_buying_2014} & differentially private peer-prediction mechanism &  & \ding{51} \\
        \cite{zhang_elicitability_2014} & - &  & \ding{51} \\
        \cite{radanovic_incentives_2014} & Divergence-based Truth Serum &  & \ding{51} \\
        \cite{radanovic_incentives_2015} & Logarithmic peer truth serum &  & \ding{51} \\
        \cite{radanovic_incentives_2016} & Peer Truth Serum (PTS) & \ding{51} & \ding{51} \\
        \cite{liu_sequential_2017} & Sequential Peer Prediction &  & \ding{51} \\
        \cite{shnayder_informed_2016} & Correlated Agreement &  & \ding{51} \\
        \cite{cvitanic_honesty_2019} & Choice-Matching &  & \ding{51}\\
        \cite{goel_personalized_2020} & Personalized PTS &  & \ding{51} \\
        \cite{kong_dominantly_2020} & Determinant based Mutual Information &  & \ding{51} \\
        \cite{schoenebeck_two_2020} & Source-differential Peer Prediction &  & \ding{51}  \\
        \cite{kamble_square_2023} & Square Root Agreement Rule &  & \ding{51} \\
        \cite{kong_dominantly_2024} & Volume mutual information &  & \ding{51} \\   
        \bottomrule
    \end{tabular}
    \label{tab:Truth_serums}
\end{table}
\end{landscape}

\newpage

\begin{landscape}
    
\begin{table}
    \centering
    \caption{Market-based mechanisms}
    \rowcolors{2}{}{gray!10}
    \begin{tabular}{*5c}
        \toprule
  
        Reference  & Mechanism name & Emp. & Nash \\
        \midrule
        \cite{baillon_bayesian_2017} & Bayesian Market & \ding{51} & \ding{51} \\
        \cite{baillon_peer_2025} & Peer betting & \ding{51} & \ding{51} \\
        \cite{ahlstrom-vij_self-resolving_2020} & Self-resolving information markets & \ding{51}&\\
        \cite{srinivasan_self-resolving_2023} & Self-resolving Prediction Markets &  & \ding{51} \\
        \bottomrule
    \end{tabular}
    \label{tab:Market_based}
\end{table}

\begin{table}
    \centering
    \caption{Output Agreement \& Proxy Scoring Mechanisms}
    \rowcolors{2}{}{gray!10}
    \begin{tabular}{*5c}
        \toprule
  
        Reference  & Mechanism name & Emp. & Nash  \\
        \midrule
        \cite{von_ahn_designing_2008} & Output Agreement & \ding{51} & \\
        \cite{witkowski_proper_2017} & Proper Proxy Scoring Rules & \ding{51} & \\
        \cite{karger_reciprocal_2021} & Reciprocal Scoring & \ding{51} &\\
        \bottomrule
    \end{tabular}
        \label{tab:Output_Agreement}
\end{table}
\end{landscape}

\newpage

\subsection*{A numerical example of the Peer-Prediction Method}

As an example, consider an academic journal: Three reviewers are asked to report whether a paper should published or not. They read the paper , i.e. receive the signal $s$, and report ($a$) to the editor (principal). 

% Graph here

Lets assume that 20\% of all paper are good and should be published and 80\% of all papers should be rejected. Furthermore, we will assume that this journal employs a set of particularly pessimistic, inaccurate and sour reviewers. The probability that a reviewer will get the signal (i.e. understand) that a paper is good, if it actually is good shall be: 

\begin{equation*}
    P(s = \text{publish} | Paper = \text{good}) = 0.4
\end{equation*}

If the paper is objectively bad, the chance of the reviewer mistaking it for a good paper is:

\begin{equation*}
    P(s = \text{publish} | Paper = \text{good}) = 0.1
\end{equation*}

The editor employs the Peer-Prediction mechanism and asks the reviewers: 

\begin{quote}
    What is the probability that a randomly chosen reference reviewer will suggest to publish the paper?    
\end{quote}

Reviewer 1 thinks the paper is good, i.e. she received the signal $s = \text{publish}$. Given that the prior probability of the paper being good (20\%) is publicly known, and that she knows of the inaccuracy of other reviewers, what should she report. Her payoff, $u$ is the squared difference between his reported probability of a peer suggesting to publish ($a_1$), and a reference peers probability of suggesting to publish ($a_i$):

\begin{equation*}
    u = - (a_1 - a_i)^2
\end{equation*}

Clearly, as the quadratic function is a proper scoring rule, it is maximized for $a_1-a_i$.  \textbf{If we assume that all other reviewers are truthful, what should reviewer 1 report?} What is the prediction that the reference reviewer $a_i$ will make? Given that we assume that the reference reviewer reports truthfully, she will base her report on her signal: 

\begin{equation*}
    P(a_i)=P(s_i)
\end{equation*}

Given that the reference reviewer has received a signal that the paper is good, the chance of the paper being objectively good is:

\begin{align*}
    P(Paper = \text{good} | s_1 = \text{publish}) &= \frac{P( s_1 = \text{publish} | Paper = \text{good} )}{P(s_1 = \text{publish})} \cdot P(Paper = \text{good}) \\
    P(Paper = \text{good} | s_1 = \text{publish}) &= \frac{ 0.4 }{0.8 \cdot 0.1+0.4 \cdot 0.2} \cdot 0.2 = 0.5
\end{align*}

The paper is objectively good with 50\% probability, given that reviewer 1 thinks it is good. Based on that knowledge, we can determine what the probability is, that the reference reviewer received the signal that the paper is good and should be published.

\begin{equation*}
    P(a_i = s_i = \text{publish}|s_1=\text{publish}) = \underbrace{0.5 \cdot 0.4}_{\text{Paper is good}} + \underbrace{0.5 \cdot 0.1}_{\text{Paper is bad}} = 0.25
\end{equation*}

Consequently, reviewer 1 should expect the reference reviewer to report a 25\% prediction, and is best off predicting 25\%. This is informative for the editor insofar that this prediction is far higher than the prior (uninformed prediction):

\begin{equation*}
    P(a_i = s_i = \text{publish}) = \underbrace{0.2 \cdot 0.4}_{\text{Paper is good}} + \underbrace{0.8 \cdot 0.1}_{\text{Paper is bad}} = 0.16
\end{equation*}

And if the reviewer 1 had gotten the signal that the paper is bad, her prediction would have been even lower, $\frac{1}{7}$ to be precise. Consequently, the editor can make an inference on whether the reviewer thought the paper is good based on her Peer-Prediction. The procedure can be adapted such that the reviewers simply directly state whether to publish or reject, and the Peer-Prediction is automatically calculated from that using the common prior and Bayes rule as shown. The paper by \textcite{miller_eliciting_2005} explains that in more detail.

\end{document}